\pdfoutput=1

\documentclass[numberedappendix]{emulateapj}
\usepackage{graphicx}

\usepackage[z,alpha]{tmsfnts}

\slugcomment{To be submitted to the Astrophysical Journal}

\def\***#1{\textbf{\textsf{*** #1 ***}}}

\shorttitle{Star formation efficiency in high-mass halos}
\shortauthors{Kravtsov A., Vikhlinin A., Mescheryakov, A.}

\begin{document}


\newcommand{\Msun}{M_{\odot}}
\renewcommand{\vec}[1]{\bmath{#1}}
\newcommand{\be}{\begin{equation}}
\newcommand{\ee}{\end{equation}}
\newcommand{\ba}{\begin{eqnarray}}
\newcommand{\ea}{\end{eqnarray}}
\newcommand{\brr}{\begin{array}}
\newcommand{\err}{\end{array}}
\newcommand{\bc}{\begin{center}}
\newcommand{\ec}{\end{center}}
\newcommand{\hm}{\,h^{-1}{\rm Mpc}}
\newcommand{\hk}{\,h^{-1}{\rm kpc}}
\newcommand{\bx}{{\bf x}}
\newcommand{\msun}{\,h^{-1}M_\odot}
\newcommand{\hMpc}{\mbox{$h^{-1}{\rmn{Mpc}}~$}}
\newcommand{\rvir}{\mbox{$R_{\rmn{vir}}$}}
\newcommand{\mvir}{\mbox{$M_{\rmn{vir}}$}}
\newcommand{\Rdc}{\mbox{$R_{\Delta_c}$}}
\newcommand{\Rtwoh}{\mbox{$R_{200}$}}
\newcommand{\Mtwoh}{\mbox{$M_{200}$}}
\newcommand{\Dc}{\mbox{$\Delta_c$}}
\newcommand{\dc}{\mbox{$\delta_c$}}
\newcommand{\Dm}{\mbox{$\Delta_{\rm m}$}}
\newcommand{\Mdc}{\mbox{$M_{\Delta_c}$}}
\newcommand{\Mfhm}{\mbox{$M_{500}$}}
\newcommand{\Mfh}{M_{500}}
\newcommand{\Rfh}{R_{500}}
\newcommand{\Mth}{M_{200}}
\newcommand{\Rth}{R_{200}}
\newcommand{\Mdm}{\mbox{$M_{\Delta_{\rm m}}$}}
\newcommand{\Mscen}{\mbox{$M_{\ast,\rm BCG}$}}
\newcommand{\Mstot}{\mbox{$M_{\ast,\rm tot}$}}
\newcommand{\Mssat}{\mbox{$M_{\ast,\rm sat}$}}
\newcommand{\fscen}{\mbox{$f_{\ast,\rm cen}$}}
\newcommand{\fssat}{\mbox{$f_{\ast,\rm sat}$}}
\newcommand{\Rd}{\mbox{$R_{\Delta}$}}
\newcommand{\rhalf}{\mbox{$r_{1/2}$}}
\newcommand{\Rhlf}{\mbox{$R_{1/2}$}}
\newcommand{\re}{\mbox{$r_{\rm e}$}}
\newcommand{\Reff}{\mbox{$R_{\rm e}$}}
\newcommand{\Md}{\mbox{$M_{\Delta}$}}
\newcommand{\Mgd}{\mbox{$M_{\rm g\Delta}$}}
\newcommand{\tmw}{\mbox{$T_{\rmn{mw}}$}}
\newcommand{\hMpcI}{\mbox{$h\,{\rmn{Mpc}}^{-1}$}}
\newcommand{\lb}{{\left<\right.}}
\newcommand{\rb}{{\left.\right>}}
\newcommand{\lum}{\,{\rm erg\,s^{-1}}}
\newcommand{\vel}{\,{\rm km\,s^{-1}}}
\newcommand{\hub}{\,{\rm km\,s^{-1}Mpc^{-1}}}
\newcommand{\lt}{$L_X$--$T$$~$}
\newcommand{\mincir}{\raise
  -2.truept\hbox{\rlap{\hbox{$\sim$}}\raise5.truept \hbox{$<$}\ }}
\newcommand{\magcir}{\raise
  -2.truept\hbox{\rlap{\hbox{$\sim$}}\raise5.truept \hbox{$>$}\ }}
\newcommand{\siml}{\raise
  -2.truept\hbox{\rlap{\hbox{$\sim$}}\raise5.truept \hbox{$<$}\ }}
\newcommand{\simg}{\raise
  -2.truept\hbox{\rlap{\hbox{$\sim$}}\raise5.truept \hbox{$>$}\ }}
\newcommand{\Mg}{M_{\rm g}}
\newcommand{\Mnl}{M_{\rm NL}}
\newcommand{\Mgas}{\mbox{$M_{\rm g}$}}
\newcommand{\fgas}{\mbox{$f_{\rm g}$}}
\newcommand{\Mstar}{\mbox{$M_{\ast}$}}
\newcommand{\Cg}{\mbox{$C_{\rm g}$}}
\newcommand{\Cs}{\mbox{$C_{\ast}$}}
\newcommand{\CT}{\mbox{$C_{\rm T}$}}
\newcommand{\ag}{\alpha_{\rm g}}
\newcommand{\as}{\alpha_{\ast}}
\newcommand{\aT}{\alpha_{\rm T}}
\newcommand{\Cgo}{\mbox{$C_{\rm g0}$}}
\newcommand{\CTo}{\mbox{$C_{\rm T0}$}}
\newcommand{\Cso}{\mbox{$C_{\ast0}$}}
\newcommand{\rhog}{\rho_{\rm g}}
\newcommand{\rhogs}{\rho_{\rm g\ast}}
\newcommand{\trhog}{\tilde{\rho}_{\rm g}}
\newcommand{\Ts}{T_{\ast}}
\newcommand{\Tx}{T_{\rm X}}
\newcommand{\Yx}{Y_{\rm X}}
\newcommand{\tT}{\tilde{T}}
\newcommand{\tphi}{\tilde{\phi}}
\newcommand{\Lx}{L_{\rm X}}
\newcommand{\Lsx}{L_{\rm Xs}}
\newcommand{\Lbol}{L_{\rm bol}}
\newcommand{\LCh}{L_{\rm Ch}}
\newcommand{\Omb}{\Omega_{\rm b}}
\newcommand{\Omm}{\Omega_{\rm m}}
\newcommand{\Oml}{\Omega_{\rm \Lambda}}
\newcommand{\Omx}{\Omega_{\rm X}}
\newcommand{\Omde}{\Omega_{\rm DE}}
\newcommand{\rhom}{\rho_{\rm m}}
\newcommand{\rhoc}{\rho_{\rm cr}}
\newcommand{\rhoco}{\rho_{\rm cr0}}

\setlength{\hbadness}{10000}


\title{Stellar mass -- halo mass relation and star formation efficiency in high-mass halos}
\author{Andrey Kravtsov\altaffilmark{1,2,3}, Alexey Vikhlinin\altaffilmark{4,5}, \& Alexander Meshscheryakov\altaffilmark{5}}
   
\altaffiltext{1}{Department of Astronomy \& Astrophysics, The University of Chicago, Chicago, IL 60637 USA, {\tt andrey@oddjob.uchicago.edu}} 
\altaffiltext{2}{Kavli Institute for Cosmological Physics, The University of Chicago, Chicago, IL 60637 USA} 
\altaffiltext{3}{Enrico Fermi Institute, The University of Chicago, Chicago, IL 60637}
\altaffiltext{4}{Harvard-Smithsonian Center for Astrophysics, 60 Garden Street,
Cambridge, MA 02138, USA}
\altaffiltext{5}{Space Research Institute (IKI), Profsoyuznaya 84 / 32, Moscow, Russia}

\begin{abstract}
We study  relation between stellar mass and halo mass for high-mass halos using a sample of galaxy clusters with accurate measurements of stellar masses from optical and ifrared data and total masses from X-ray observations.  We find that stellar mass of the brightest cluster galaxies (BCGs) scales as $M_{*,\rm BCG}\propto M_{500}^{\alpha_{\rm BCG}}$ with the best fit slope of $\alpha_{\rm BCG}\approx 0.4\pm 0.1$. We measure scatter of $M_{*,\rm BCG}$ at a fixed $M_{500}$ of $\approx 0.2$ dex.   We show that stellar mass--halo mass relations from abundance matching or halo modelling reported in recent studies underestimate masses of BCGs by a factor of $\sim 2-4$. We argue that this is because these studies used stellar mass functions (SMF) based on photometry that severely underestimates the outer surface brightness profiles of massive galaxies. We show that $M_{\ast}-M$ relation derived using abundance matching with the recent SMF calibration by Bernardi et al. (2013) based on improved photometry is in a much better agreement with the relation we derive via direct calibration for observed clusters. The total stellar mass of galaxies correlates with total mass $M_{500}$ with the slope of $\approx 0.6\pm 0.1$ and scatter of $0.1$ dex. This indicates that efficiency with which baryons are converted into stars decreases with increasing cluster mass.
The low scatter  is due to large contribution
of satellite galaxies: the stellar mass in  satellite galaxies correlates with $M_{500}$ with scatter of $\approx 0.1$ dex and best fit slope of $\alpha_{\rm sat}\approx 0.8\pm 0.1$. 
We show that for a fixed choice of the initial mass function (IMF)  total stellar fraction in clusters is only a factor of $\sim 3-5$ lower than the peak stellar fraction reached in $M\approx 10^{12}\rm\ M_{\odot}$ halos. The difference is only a factor of $\sim 1.5-3$ if the IMF becomes progressively more bottom heavy with increasing mass in early type galaxies, as indicated by recent observational analyses.  This means that the overall efficiency of star formation in massive halos is only moderately suppressed compared to $L_{\ast}$ galaxies and is considerably less suppressed than previously thought. The larger normalization and slope of the $M_{\ast}-M$ relation derived in this study shows that feedback and associated suppression of star formation in massive halos should be weaker than assumed in most of the current semi-analytic models and simulations.  

\end{abstract}


\section{Introduction}
\label{sec:intro}

In hierarchical Cold Dark Matter (CDM) models of structure formation,
galaxies are thought to form via dissipative processes within
potential wells of dark matter-dominated halos
\citep{white_rees78,fall_efstathiou80,blumenthal_etal82,blumenthal_etal84}. Cooling
leads to condensation of baryons towards the center of their parent
halo \citep[e.g.,][]{rees_ostriker77} where they reach conditions
suitable for star formation.  Understanding efficiency with which
halos convert their baryons into stars is one of the central problems
in modelling galaxy formation. Results of abundance matching of halo mass
function to the stellar mass function of galaxies showed that this
efficiency peaks at halo masses of $M\approx 10^{12}\,\rm M_{\odot}$ but
is quickly decreasing at both smaller and larger masses
\citep[e.g.,][]{conroy_wechsler09,guo_etal10}. Remarkably, this peak
mass is almost independent of redshift
\citep{yang_etal12,behroozi_etal13b}.  Even at
the peak of efficiency, observed masses of cold gas and stars in
galaxies corresponds to only $\approx 20-30\%$ of the mass of baryons
which should have been potentially available in $M\approx 10^{12}\,\rm
M_{\odot}$ halos. The implied low star formation efficiencies 
 are widely believed to be due to strong feedback
processes, which suppress star formation and blow out a large fraction of
gas out of host halos.

In particular, it is commonly acknowledged that the AGN feedback plays a key
role in shaping properties of baryons in massive halos and in setting
their star formation efficiency \citep[e.g.,][and
\citealt{mcnamara_nulsen07} for a recent
review]{silk_rees98}. Semi-analytical models showed that without AGN
feedback one cannot reproduce the high mass end of stellar mass
function \citep[e.g.,][]{benson_etal03,croton_etal06}. Simulations
carried out over the past decade clearly showed that without strong
AGN feedback stellar fractions in groups and clusters are
overestimated by a factor of $\sim 2-3$ compared to observations
\citep[see][for a recent review]{kravtsov_borgani12}. Inclusion of
strong AGN feedback in simulations using phenomenological recipes indeed lowers
stellar fractions by a significant factor
\citep[e.g.,][]{mccarthy_etal10,mccarthy_etal11,planelles_etal13,martizzi_etal12,martizzi_etal13,ragone_figueroa_etal13}.

Although these results are encouraging, they are not yet
conclusive. Details of how AGN feedback affects evolution of massive
galaxies and properties of their gaseous halos are not yet fully
understood.  Implementations of AGN feedback in semi-analytic models and cosmological 
simulations are thus associated with significant systematic uncertainties, which means that
results of model calculations need to be thoroughly tested against observations. Group and
cluster scale halos provide a unique way of testing models because
stellar masses of galaxies, halo gas content, and halo mass all can be
estimated from observations fairly reliably.
 
In this paper, we use a sample of nine clusters with carefully
measured stellar masses and total masses to calibrate
efficiency of star formation and stellar mass--halo mass relation in
massive halos. We combine this sample with 12 clusters with similar
data from a recent study of \citet{gonzalez_etal13} to improve the
statistics in calibrations of trends. Although stellar content of
clusters was investigated in a number of recent studies
\citep{gonzalez_etal07,gonzalez_etal13,lagana_etal13,budzynski_etal14},
most of these works focused on the total stellar and hot gas content
only to estimate total baryon budget of clusters, rather than on
stellar mass--halo mass relation of galaxies specifically. In this
study we consider both the total stellar mass within halos and masses
of the BCG and satellite galaxies separately, because transition
from galaxy-sized halos to groups and clusters is accompanied by
increasing fraction of stellar mass in satellite galaxies. 

In addition, the fraction of stellar mass in the outer regions around central galaxies also increases with increasing halo
mass. In galaxy clusters the outer regions of the brightest cluster
galaxy are often called the intra-cluster light (ICL) and are estimated 
to contain significant fraction of the BCG mass \citep{gonzalez_etal05} and up to $\sim
20-40\%$ of the total stellar mass within the virial radius
\citep{zibetti_etal05,gonzalez_etal07,skibba_etal07,gonzalez_etal13,mcgee_balogh10}. Thus,
the outer profiles of BCGs need to be properly accounted for when considering stellar content
of clusters. Profiles at large radii can, however, be easily missed or under-estimated in
observations due to their low surface brightness and challenges in
estimating background around large massive galaxies in
crowded environments. 

In this study we use stellar masses measured in re-analysis of the SDSS 
data with a careful treatment of background, which allows us to 
recover the low surface brightness outer profiles of galaxies out to $\approx 100-300$ kpc. 
 We use these measurements along with X-ray estimates
of total cluster mass to calibrate 
stellar mass-halo mass relation and stellar fractions of high mass halos. 

The paper is organized as follows. We present our stellar surface density and mass
measurements in Section \ref{sec:data}. We show that stellar masses of BCGs
in nearby clusters based on the standard SDSS photometry are
underestimated by a factor of $\sim 2-4$. At the
same time, our measurements are in good agreement with luminosities
derived by \citet{bernardi_etal13} using re-analysis of the SDSS data
with improved estimate of background. We show that stellar mass of BCGs
continues to increase at the outermost
measured radii in most clusters.  

We present calibrations of stellar mass--halo mass relation for a
combined sample of 21 clusters in Section \ref{sec:msmh}. We compare our
observational calibrations with expectations from the halo abundance
ansatz and show that they are in reasonable agreement if the recent calibration of the SMF by \citet{bernardi_etal13} is used in the latter. In Section \ref{sec:bcgpro} we  show that BCG {\it size}  correlates with cluster halo virial radius and extends the corresponding correlation
of smaller mass galaxies. 

We discuss interpretation of our stellar mass--halo mass relation in
terms of efficiency of star formation in halos of different mass in
Section \ref{sec:fstar}. We show that inclusion of the outer regions of BCGs in stellar mass measurement
considerably increases the stellar mass 
and star formation efficiency estimates. We also point out that if
the initial mass function of stars becomes substantially more
bottom-heavy in massive early type galaxies, as indicated by a number
of recent observational analyses, the star formation efficiency in
cluster halos is only a factor of $\sim 2-3$ lower than the efficiency
of $M\approx 10^{12}\ \rm M_{\odot}$ halos. We discuss our results in Section \ref{sec:disc}, comparing them 
with results of previous studies of stellar fraction as a function of
mass in Section \ref{sec:leaucomp}. We discuss implications of our results for the strength of feedback in semi-analytic models and cosmological
simulations in Section \ref{sec:impl}.
We summarize our results and conclusions in Section \ref{sec:conc}.

Unless otherwise noted, throughout this paper we assume a flat $\Lambda$CDM model with parameters $\Omega_{\rm m}=1-\Omega_\Lambda=0.27$, $\Omega_{\rm b}=0.0469$, $h=H_0/(100\rm\,km\,s^{-1}Mpc^{-1})=0.7$, $\sigma_8=0.82$ and $n_{\rm s}=0.95$ compatible with combined constraints from WMAP, BAO, SNe, and cluster abundance \citep{vikhlinin_etal09b,komatsu_etal11,hinshaw_etal13}. Total masses are defined within radius enclosing a particular overdensity ($500$ or $200$) with respect to the critical density at redshift of observation, which is indicated by a corresponding subscript ($M_{500}$ or $M_{200}$). 
Throughout the paper we assume the \citet{chabrier03} IMF in calculation of stellar masses, except in \S~\ref{sec:imf} where we explore sensitivity of stellar fractions to IMF variation. 

\section{Stellar masses of group and cluster galaxies}
\label{sec:data}

To calibrate the stellar mass -- halo mass relation on the high mass
end, we use two sets of measurements for a combined sample of 21
clusters with $z\lesssim 0.1$ spanning the total mass range of
$\Mfh\approx (0.5\div 15)\times 10^{14}\ \rm M_{\odot}$. The clusters
in the sample have individual measurements of stellar masses, which
include  measurements of low surface brightness outer regions of massive galaxies, and high-quality X-ray data, which are used to
get accurate estimates of the total masses of individual
clusters. 

The first set is taken from a recent study by \citet[][hereafter
G13]{gonzalez_etal13}, who used re-analysis of the $I$ band photometry
for a subset of 12 clusters from \citet{gonzalez_etal07} complemented
with the XMM-{\sl Newton} X-ray data.  The total mass, $\Mfh$, for
these clusters is estimated using X-ray temperature obtained by the
spectral fit within the radial range $(0.15-0.5)\, \Rfh$ via an
iterative procedure and a fitting formula of
\citet{vikhlinin_etal09a}. The total masses derived by G13 are in good
agreement with the masses derived by \citet{vikhlinin_etal06} for the
same clusters, but are slightly offset low by an average of $\approx
8\%$. This offset is independent of mass and reflects a small
systematic difference between \emph{Chandra} and \emph{XMM-Newton}
cluster temperature measurements \citep{vikhlinin_etal05}. Its
magnitude is too low to affect our results and conclusions.  The
total $I$-band luminosities of galaxies were derived from drift-scan
observations at the Las Campanas 1\,m Swope telescope by carefully
modelling profiles of galaxies to large radii and low surface brightness, as detailed in \citet{gonzalez_etal05}. 

The second set of measurements is new and was derived using the SDSS
data for nine nearby ($z<0.1$) clusters, all of which fall within the
SDSS footprint and have high-quality {\sl Chandra} X-ray data. The
sample was chosen to span a wide range of total masses (see
Table~\ref{tab:sdsscl}) and to avoid highly disturbed systems in which
we cannot assume spherical symmetry. The total masses, $\Mfh$, were
derived using robust, low-scatter mass proxy $\Yx$
\citep{kravtsov_etal06} derived from the {\sl Chandra} observations, as desribed in \citet{vikhlinin_etal09a}, except for
Abell\,1991, MKW\,4, RXJ\,1159+5531. For the latter three clusters the
gas and total masses were estimated from \emph{Chandra} X-ray data
using the identical analysis. We describe the procedure used to derive
the stellar masses next.

\subsection{Stellar mass measurements for nine clusters using the SDSS data}
\label{sec:sdssbcgmass}

The stellar masses were measured using raw images and corresponding
calibration data from the Sloan Digital Sky Survey Release 8 in $g$,
$r$, and $i$ bands\footnote{{\tt http://www.sdss.org/dr8/}} in the
$\approx 38.6^\prime\times 42^\prime$\footnote{Corresponding to physical areas of $\approx 2\times 2 R_{500}$ for most clusters, and $\approx 1.6\times 1.6 R_{500}$ for MKW4, $\approx 3.4\times 3.8 R_{500}$ for Abell 1991, $\approx 6.0\times 6.8 R_{500}$ RXJ 1159}, aligned mosaic fields centered on the
brightest galaxy in each cluster. The mosaic images in each band were
constructed by combining together four fields of the same SDSS drift
scan (run) that contain cluster BCG into $2048\times 6933$ pixel
strips and then gluing together three neighboring strips. These three
strips came from different runs and thus have patterns of background
non-uniformities. The strongest background non-uniformities in the SDSS
images are in the scan direction ($y$). Fortunately, the individual
SDSS scans overlap, and therefore the difference in the background
patterns in the scan direction can be eliminated. We achieved this by
fitting a low-order polynomial to the difference in the 1D profiles in
the overlapping region of the neighboring strips, and subtracted that
fit from the strips adjacent to that containing the brightest cluster 
galaxy. After the strips were joined, we subtracted the global
background profile from the whole mosaic image. The global background
was estimated in a region of width equal to $300-500$ pixels and length
equal to the entire mosaic in the $y$-direction, chosen in an area
without bright stars or extended objects and at the distance $> 1500$
pixels ($\gtrsim 250-800$ kpc) from the BCG center (known sources were
also masked out). This region was used to extract a binned
one-dimensional profile in the $y$ direction, which was fit by a
low-order polynomial. The background profile obtained in this way was
then subtracted from the entire $6000\times 6000$ pixel strip.

The images were de-blended from nearby stars and galaxies and
appropriate mask regions were created in order to subtract
contribution from remaining foreground/background objects (stars,
galaxies and artefacts).  Radial surface brightness in different bands
and $(g-r)$ color profiles of the BCGs were then extracted from
the de-blended and masked mosaic images. 

In order to take into account possible additional residual variations
of the background across the field, we fitted the quadratic model for
residual background, $f(R)=C_0+C_2R^2$, where $R$ is the radial
distance from the BCG center, during the profile extraction. Note that
all linear and bilinear background components are averaged out in the
azimuthal profile, and therefore thus defined $f(R)$ represents the
lowest-order non-constant local background adjustment to the galaxy
light profile. The $C_{0}$ and $C_{2}$ coefficients are determined by
fitting the data in the at $R\gtrsim 600-1400$ pixels ($\gtrsim 4-9'$), where the
contribution of the BCG light is neglibible. The
best-fit $f(R)$ model is then interpolated to smaller radii, where the
BCG profile measurements are typically done, $\lesssim 3.5'$. We
evaluated the residual background uncertainties as a function of
radius by applying this procedure to the SDSS mosaic images of the same
size centered on random regions in the SDSS data that do not contain
known nearby clusters. This test shows that the residual $r$-band background
uncertainties in our BCG profiles are 28.4~mag~arcsec$^{-2}$ at
$R=2'$, and 28.8~mag~arcsec$^{-2}$ at $4'$. The image mosaic and background estimation procedure described above was done for each BCG galaxy of the nine
clusters in our sample and for second brightest galaxy in Abell 2142 and MKW3s. For satellite galaxies galaxy surface brightness profile plus a constant background, with corresponding masks and de-blending when needed, was measured within fields of 400 pix size centered on each galaxy. In a few difficult cases (out of several thousand total galaxies measured) the profiles were measured using extended regions of $\approx 600-1000$ pixels.

After subtracting the best-fit local background model, the BCG
brightness profiles in $g$, $r$, and $i$ bands were obtained for all
BCGs in our cluster sample. 
Conversion from counts on the SDSS images
to apparent AB magnitudes was done following the standard
procedure\footnote{See {\tt
    http://www.sdss.org/dr7/algorithms/fluxcal.html}}. The apparent
magnitudes were then converted to the absolute magnitudes using $m-M =
A + K(z) + EC(z) + DM(z,M,L,h)$, where $A$, $K(z)$, $EC(z)$, $DM$ are
the Galactic absorption correction from the \citet{schlegel_etal98}
dust map individual for each cluster, $K$-correction, evolutionary
correction and distance modulus. For galaxies with known redshifts the
$K$-corrections were derived individually by using kcorrect (v 4.1.4)
code \cite[][]{blanton_roweis07} and $u$, $g$, $r$, $i$, $z$-band
fluxes derived from {\tt model} magnitudes in SDSS catalog, while for
the rest of galaxies we used the median mean $K$-correction of galaxies
with redshifts. It should be noted that the $K$-corrections at
redshifts considered here are small. The evolutionary corrections were
estimated using approximate relation of \citet{bell_etal03}: $EC(g, r,
i)\approx (-1.6, -1.3, -1.1)z$.  To convert absolute magnitudes of
galaxies into luminosities in solar units, we used absolute
magnitudes of the Sun $M_g=5.15$, $M_r=4.67$, $M_i=4.56$ \citep{bell_etal03}.

The surface brightness profiles of BCG galaxies were fit with the triple
S\'ersic profile within the radius $R_{\rm out}$, where error in the
profile due to background uncertainty reaches $33\%$. We find that 
three components provide a far more accurate fit than a single S\`ersic fit. This 
was also recently found to be true for early type galaxies in general \citep{huang_etal13a}. The total magnitude of galaxy is estimated by extrapolating the model profile
to infinity. 

In Abell 2142 a special care needs to be taken in deriving the mass of the BCG, because the cluster contains a second bright galaxy of somewhat smaller luminosity near the BCG. We have de-blended the second bright galaxy by first deriving the profile of the BCG with the second brightest galaxy masked with a circular aperture of radius  $53^{\prime\prime}\approx 87$ kpc.
The resulting surface brightness profile was fit with  the triple
S\'ersic profile. This profile was then was subtracted from the image and the circular region of radius $75^{\prime\prime}\approx 123$ kpc around the center of the BCG was masked out. The profile of the second brightest galaxy was then derived from the resulting image. In this procedure, any stellar envelope of the second brightest galaxy at large radii will be assigned to the BCG. To test for potential effect of such contribution, we have also recalculated BCG mass decreasing the stellar mass at $r>170$ kpc by a factor of two. We have checked that all of the scaling relations measured below do not change if such reduced BCG mass is used for Abell 2142. MKW3s cluster has also a bright galaxy close to the BCG. In this case, the BCG mass was computed in a similar way to that of Abell 2142, but with preliminary profile of the second galaxy measured at small radii subtracted before measurement of the BCG profile. The BCG profile was then measured with a circle of radius  $58^{\prime\prime}\approx 50$ kpc masked around the second brightest galaxy. In this case, the second brightest galaxy has significantly lower luminosity than BCG and such procedure is more appropriate.
The measured BCG profile was then subtracted from the image and profiles of other galaxies were measured with circle of radius $190^{\prime\prime} = 164$ kpc around BCG masked out.

\begin{figure}[t]
\vspace{-1cm}
\begin{center}
\includegraphics[scale=0.48]{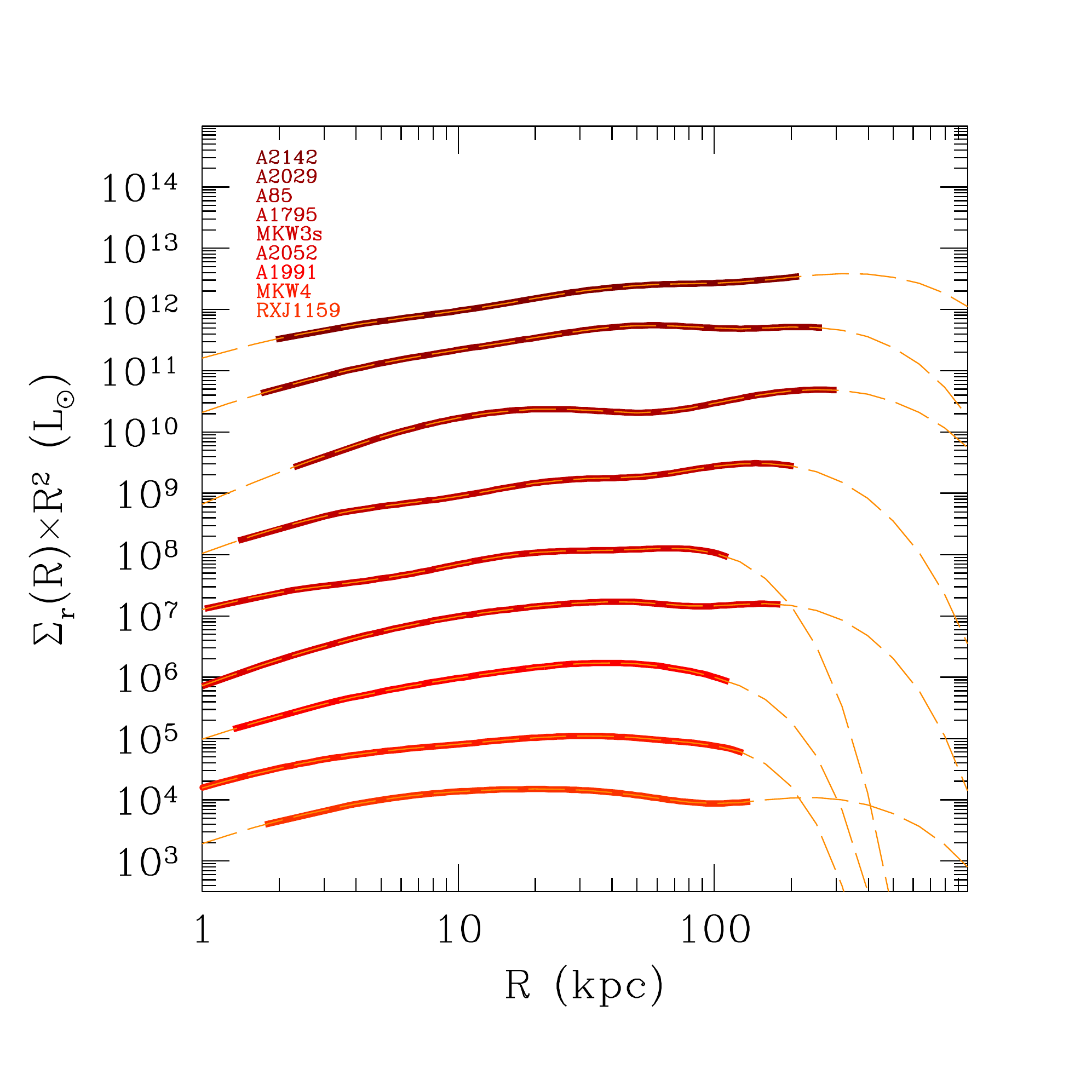}
\vspace{-1cm}
\caption{Profiles of $r$-band luminosity surface density of BCGs in nine clusters analyzed in this study, as indicated in the legend. Luminosities are multiplied by $R^2$ and profiles are displaced by a dex with respect to each other for clarity (the third from the top profile of BCG in Abell 85 is not shifted). The thick lines are the actual measured profiles shown out to the radius $R_{\rm out}$ (see Table~\ref{tab:sdsscl}) where background errors reach $33\%$, while the long dashed thin lines show the best three S\'ersic component fit.}
\label{fig:bcgpror2}
\end{center}
\end{figure}

The results are reported in Table\,\ref{tab:sdsscl}, while Figure~\ref{fig:bcgpror2} shows the surface density profiles of $r$-band luminosity for BCGs in nine clusters along with the  three S\'ersic 
component best fit. The figure shows that luminosity increases at the outermost measured radius, $R_{\rm out}$,  in all BCGs and shows how much of the total luminosity is contributed by extrapolation of the fit beyond $R_{\rm out}$. In the following analysis, we will be using stellar masses measured by extrapolation of the triple-S\'ersic fit to their profiles within $R_{\rm out}$ to infinity. Such extrapolation is a standard practice in measurements of galaxy luminosity, but is of course somewhat uncertain as the actual profile at large radii may deviate from the 
extrapolated one. As a conservative estimate for how much the stellar mass could underestimated, we have also computed total BCG stellar masses by extrapolating their profiles at $R>R_{\rm out}$ as $\Sigma(R)=\Sigma(R_{\rm out})(R/R_{\rm out})^{-2}$, i.e. profile approximately describing $\Sigma(R)$ of most galaxies at $R\approx R_{\rm out}$ out to $R_{500}$. Such masses are on average $\approx 30\%$ larger than the $M_{\ast,\rm BCG}$ masses in Table\,\ref{tab:sdsscl}. We have checked that using such masses simply increases normalization of the best fit $M_{\ast,\rm BCG}-M_{500}$ relation presented in Section~\ref{sec:mscmh} (Table 2 and Fig. \ref{fig:msmhhigh}) by $\approx 30\%$ ($\approx 1.5\sigma$) but does not change slope and scatter significantly. Therefore, even with such maximum correction does not influence our results and conclusions appreciably.

\subsection{Comparison with other luminosity measurements}
\label{sec:lcomp}

\begin{figure}[t]
\begin{center}
\vspace{-1cm}
\includegraphics[scale=0.48]{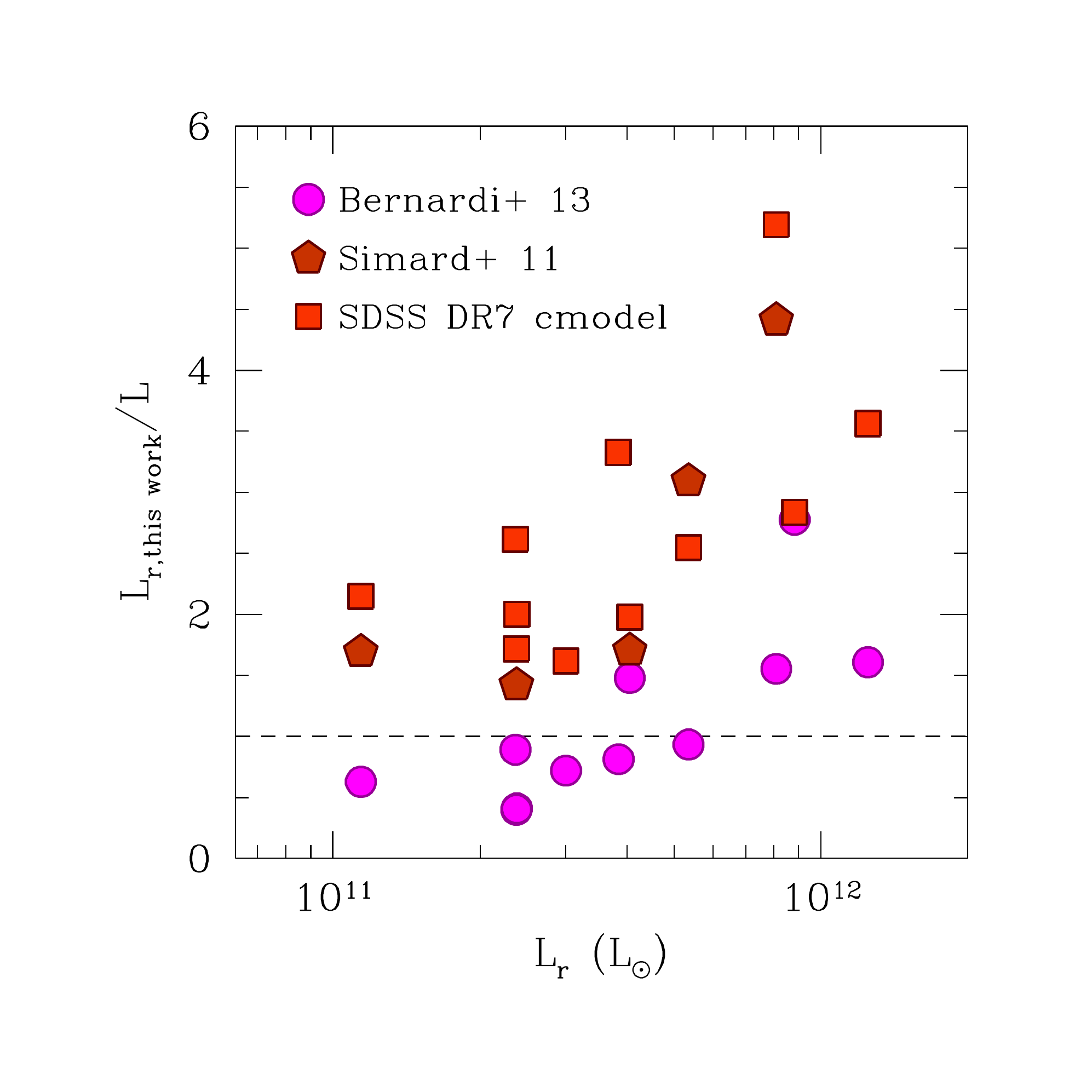}
\vspace{-1cm}
\caption{Ratio of luminosities of the BCG galaxies derived in this study to the luminosities of these galaxies in the SDSS DR7 catalog, catalog of \protect\citet{simard_etal11} and luminosities derived from PyMorph fits of \citet{bernardi_etal13}. }
\label{fig:bcgcompl}
\end{center}

\end{figure}

Figure~\ref{fig:bcgcompl} compares luminosities derived in our
analysis described above with the luminosities calculated from the
\texttt{cmodel} magnitudes from the SDSS DR7 pipeline,\footnote{We compare to the DR7 photometry because this data release was used for recent stellar mass function measurements. However, we have checked that conclusions do not change if use photometry from the DR8 release.} and with the
re-analyses of the photometry by \citet{simard_etal11} and
\citet{bernardi_etal13}. 
The luminosities for the latter two studies
were calculated from the S\'ersic fit to the $r$-band
data.\footnote{Photometry results from the study of
  \citet{bernardi_etal13} for our galaxies was kindly communicated to
  us by A. Meert and M. Bernardi.} The SDSS \texttt{cmodel} luminosities were
computed using {\tt deV} and {\tt exp} magnitudes for the same
galaxies from the SDSS DR7 database using equation 1 in
\citet{bernardi_etal10} and applying sky-subtraction correction given
by equation 4 in the same paper.

Figure~\ref{fig:bcgcompl} shows that luminosities derived in our analysis are in reasonable agreement with luminosities of \citet{bernardi_etal13}, although the scatter between the two measurements is substantial. Luminosities from the cmodel magnitudes are biased low by a factor of $\sim 2-4$ and there is an indication that the bias increases with increasing luminosity. Luminosities derived from the absolute magnitudes reported by \citet{simard_etal11} for five of our galaxies that can be found in their catalog are also biased low, although the magnitude of the bias is somewhat smaller than for the cmodel luminosities. This indicates that the bias is not due to a particular choice of the functional form used to fit surface brightness profiles. Note that similar biases were reported by \citet{bernardi_etal13}, who show that cmodel and \citet{simard_etal11} absolute magnitudes are biased high by $\approx 0.7$ and $\approx 0.5$ magnitudes, respectively, at the absolute magnitudes characteristic of the cluster BCGs ($M_r\sim -24\div -26$). These differences correspond to factors of $\approx 1.6$ and $\approx 1.9$ in luminosity difference and are comparable to the results shown in Figure~\ref{fig:bcgcompl} for lower luminosity BCGs. Our results indicate that bias for the most luminous BCGs is even larger and reaches factors of $\approx 3-4$.

\begin{figure}[t]
\vspace{-1cm}
\includegraphics[scale=0.48]{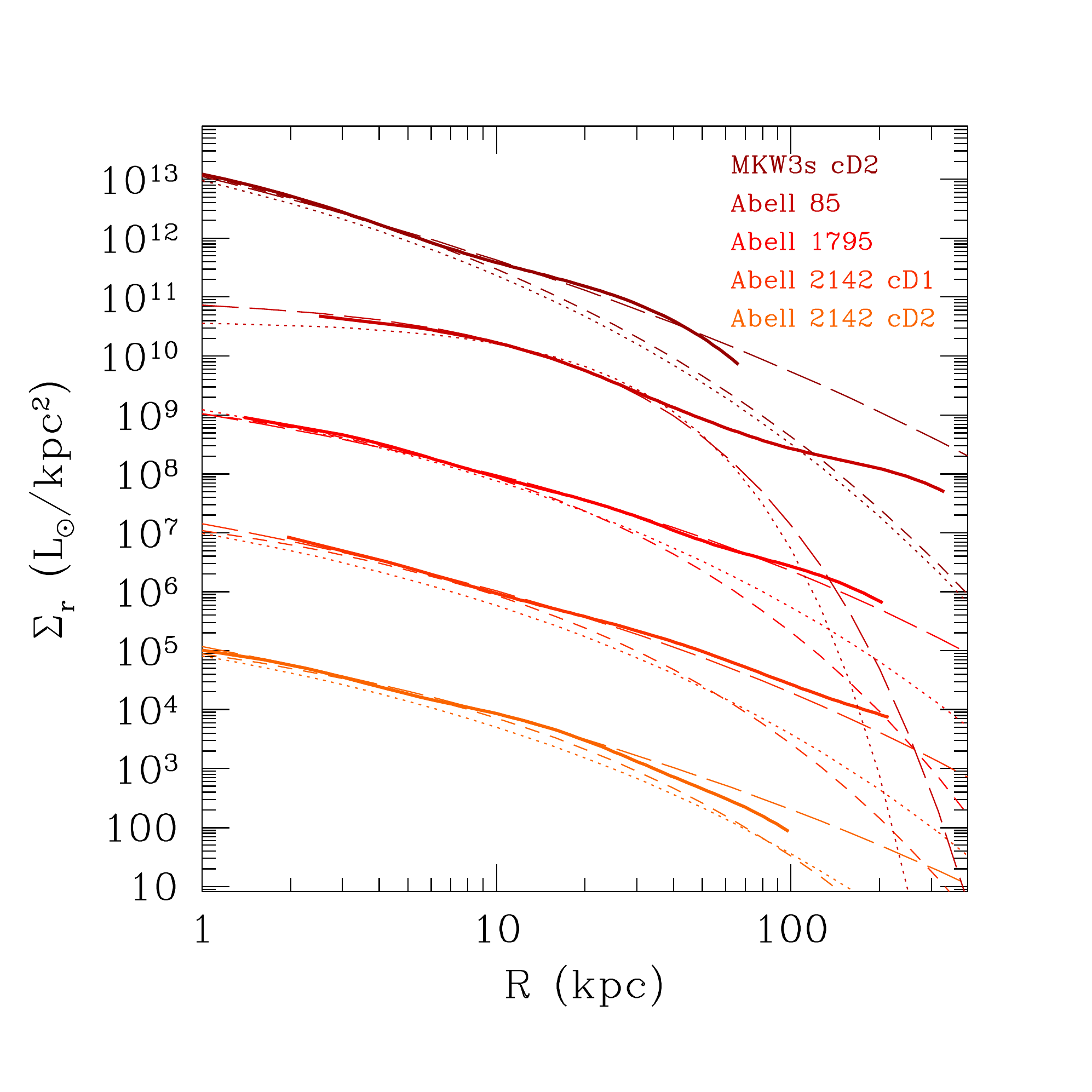}
\vspace{-1cm}
\caption{Comparison of the $r$-band luminosity surface density profiles for five cluster galaxies derived in this study ({\it thick solid lines}), SDSS DR7 pipeline ({\it dotted lines}),  by the 2D photometric S\'ersic fits of \protect\citet[][{\it short dashed lines}]{simard_etal11} and \protect\citet[][{\it long dashed lines}]{bernardi_etal13}. 
The profiles derived in this study are shown to the outer radius, where error in the surface density profile has reached 30\% due to uncertainties in the background subtraction.
Profile for the BCG in Abell 1795 is shown with proper normalization, while normalizations of other profiles is rescaled by two orders of magnitude from each other for clarity. We show galaxies present in the SDSS DR 7 catalog and for which photometry from \protect\citet{bernardi_etal13} and \protect\citet{simard_etal11} was available. Note that the BCG of Abell 85 is absent from the catalog of \protect\citet{simard_etal11}. For cluster Abell 2142 we show profiles of the two brightest cluster galaxies. }
\label{fig:bcgcompspro}
\end{figure}

The origin of the bias is clear from the comparison of the surface brightness profiles shown in Figure~\ref{fig:bcgcompspro}. In this figure we show the $r$-band luminosity surface density profiles for five galaxies for which there is corresponding photometry in the SDSS DR7 catalog along with the results of the S\'ersic fits using the {\tt PyMorph} method by \citet{bernardi_etal13}, and 2D S\'ersic fits from the catalog of \citet{simard_etal11}. In the latter photometric redshifts of the galaxies were used to compute luminosities and radii for some of the galaxies. We have rescaled these to the proper spectroscopic redshifts used to derive luminosities in our study. Note that BCG of Abell 85 is absent from the \citet{simard_etal11} catalog.

The figure shows that standard SDSS photometry underestimates the luminosity density somewhat at small radii for all galaxies and severely underestimates the luminosity density at large radii for all galaxies. The best fit profiles of \citet{simard_etal11} are not biased at small radii, but have equally large error at large radii. 
As shown by \citet{lauer_etal07} and \citet{bernardi_etal07},  the main source of the bias is over-subtraction of background \citep[see also][]{he_etal13} that is estimated at radii that are too small for such luminous, extended galaxies in general, and BCGs in particular. Recently, \citet{bernardi_etal13} and \citet{he_etal13} showed that this bias increases with increasing galaxy luminosity and significantly affects  galaxies on the exponential part of the luminosity function ($L>L_*$). 
\citet{bernardi_etal13} correct for this problem in their analysis and their best fit S\'ersic profiles are in a much better agreement with the profiles derived in this study at large radii. The only exception is BCG in Abell 85, which has a very shallow outer profile
that single S\'ersic component does not fit well. 

Overall our results confirm the bias underestimating luminosities by the SDSS photometric pipeline and the fact that bias increases with increasing galaxy luminosity. Comparisons indicate that luminosities derived using the single S\'ersic fits with proper handling of sky background subtraction by \citet{bernardi_etal13} are not significantly biased with respect to our photometric analysis.

\subsection{Measuring luminosities of satellite galaxies}

In addition to the light of the BCG, we measure luminosities associated 
with other cluster galaxies as follows. We first construct 
a complete list of galaxies up to a
cluster-centric radius of $\approx 2-4^{\circ}$ ($\approx 10-20\times R_{500}$)  from the 
SDSS photometric catalog using only objects detected with $\gtrsim 5\sigma$
significance in $g$, $r$, and $i$ bands and which are classified as 
{\tt GALAXY} in at least two of the three
photometric bands. We have rejected all objects with saturated pixels and
which are flagged as {\tt BRIGHT}.  

We use the selection criteria of \citet{yasuda_etal01} to select only objects flagged in the SDSS catalog as {\tt PRIMARY}, which avoids multiple detections in the overlap between adjacent scan lines in two strips and
between adjacent frames. Furthermore, we use only isolated objects, child objects and {\tt NODEBLEND} flagged objects in constructing the galaxy
sample. 

The object list is cleaned to remove 
background and foreground galaxies using tiling data from rhe SDSS spectroscopic catalog, which allows us to substantially reduce the statistical background for galaxies brighter than $r=17.5$ mag.  We use the USNO B1 catalog to identify very bright stars in the field and mask out regions in which light from star affects photometry of galaxies.

In addition, regions containing visible prominent density
peaks in galaxy distribution in the neighbourhood of each cluster were masked.
Furthermore, we use images from the ROSAT All Sky Survey to identify nearby clusters and
 mask out regions of diffuse X-ray sources from our analysis. This avoids any effect 
projecting substructure may have on the radial distribution of cluster galaxies.

As we discussed above, for bright galaxies the luminosity may be significantly underestimated if background is overestimated. Although the problem is most
severe for the brightest galaxies, it can also affect luminosities of lower mass
systems. We have compared total luminosities and colors obtained from the magnitudes in the SDSS catalog with values from our own re-analysis of photometry using profile fitting procedure similar
to that described above for the BCG galaxies, except for lower luminosity galaxies we fitted two S\'ersic components rather than three. The total galaxy luminosities in our re-analysis were obtained by extrapolating the best fit two-S\'ersic profiles to infinity. We performed such re-analysis for all galaxies
within $6R_{500}$ of the cluster center and with $L_r\gtrsim 1.5-10\times 10^9\,\rm L_{\odot}$ (the limit varied from cluster to cluster). The luminosities derived in this way are generally higher than luminosities from the SDSS photometry for galaxies with with absolute magnitudes of $M_{\rm r}\gtrsim -21$ with difference increasing with increasing luminosity, consistent with results of \citet{bernardi_etal13}. We have fit a low-order polynomial for the luminosity dependence of the difference in magnitudes and applied this correction
for low-luminosity galaxies below the threshold for our photometric re-analysis. 

 To estimate galaxy colors
we integrated best fit surface brightness profiles models within aperture
$R_{\rm col} = \max(R_{\rm eff} , 5\ {\rm kpc})$, 
where $R_{\rm eff}$ was calculated from the $R_{\rm eff}-L_{\rm r}$ relation of elliptical galaxies \citep[e.g.,][]{desroches_etal07}. We find colors that are consistent with colors from the SDSS catalog. Given luminosities and colors, stellar masses are estimated using relation between $M_\ast/L$ and color, as described in the next section. 

Using the measured stellar masses we construct stellar mass function (SMF) of cluster galaxies.  For each cluster, we estimate the number of galaxies in a given luminosity bin $j$ as $N_{\rm j}=N_{\rm a,j}-N_{\rm b,j}S_{\rm a}/S_{\rm b}$, where $N_{\rm a,j}$ is the number of galaxies (cluster$+$field) within the cluster region, $N_{\rm b,j}$ is the number of galaxies in the background region, and $S_{\rm a}$ and $S_{\rm b}$ are the areas of cluster and
field region, respectively. We choose the field region to estimate the background at $R\approx 5-6R_{500}$ from the cluster X-ray peak with area such that $S_{\rm a}/S_{\rm b}\ll 1$, while cluster aperture is chosen to be a circle of radius $0.7R_{500}$ centered on the X-ray peak.

The SMF of cluster galaxies shows a prominent break and starts to decrease at stellar mass $M_\ast\lesssim 1-5\times 10^9\ \rm M_{\odot}$, which corresponds to apparent magnitude of $m_{\rm r}\approx 20.5$ (this break varies within this range from cluster to cluster in terms of stellar mass).
We thus adopt completeness limit, $M_{\rm lim}$, for the stellar mass equal to twice the value of the break estimated for each cluster and fit Schechter form to the SMF for galaxies with $M_\ast>M_{\rm lim}$, excluding the BCG. The SMF normalized by total mass, $d(N/M_{500})/dM_\ast$, of eight clusters, except MKW 4, can be described by the same Schechter function, $d(N/M_{500})/dM_\ast\propto (M_\ast/M_{\ast,\rm ch})^{\alpha}\exp(-\left[M/M_{\ast,\rm ch}\right])$ ($\chi^2<2.7$) with normalization of $6.15\times 10^{-25}$, $\alpha=-1.09$ and $M_{\rm \ast,b}=1.23\times 10^{11}\rm M_{\odot}$. Extrapolating the best fit functional
fit to SMF below $M_{\rm lim}$, we estimate contribution of galaxies with $M_{\ast}<M_{\ast\rm lim}$ to be $\approx 10-20$\% for eight of the clusters in our sample.  MKW 4 is the
nearest cluster in our sample and has the absolute magnitude limit two  magnitudes 
fainter than for other clusters. For this cluster we fit SMF of cluster galaxies with a double Schechter fit fixing the characteristic mass of bright galaxies to the above value for the rest of the clusters, but varying all other parameters. We derive best fit slope of the bright-end Schechter function of $\alpha_{\rm b}=-0.66$; at the faint end the best fit slope and characteristic mass are $\alpha_{\rm f}=-1.82$ and $M_{\ast,\rm f}=9.05\times 10^9\rm\ M_{\odot}$. 
Based on this fit, contribution of galaxies with $M_{\ast}<M_{\ast\rm lim}$ for MKW 4 is to be $\approx 26$\% for the nine clusters in our sample. If SMFs in the other eight clusters have a similar upturn at small masses, we may be missing up to $\approx 25\%$ of stellar light in the estimate of the total stellar mass in satellite galaxies.

To measure projected (2D) radial distribution of galaxies we use only the galaxies
with $M_{\ast} > M_{\ast\rm lim}$. We de-project the 2D profile by forward fitting a 3D $(\alpha,\beta,\gamma)$ profile model, $\rho(r)=A(r/R_{\rm s})^{-\alpha}(1+\left[r/R_{\rm s}\right]^{\beta})^{-(\gamma-\alpha)/\beta}$ \citep{lauer_etal95,zhao96} plus constant background, $B$, to the projected profile, where $\alpha$ controls the inner slope, $\gamma$ controls the outer slope, and $\beta$ controls how fast the slope changes around $R=R_{\rm s}$.  We fix $\alpha=0$, $\beta=3$ in out fits and fit for normalization $A$, $R_{\rm s}$, $\gamma$, and $B$.  The best fit profiles are used to estimate stellar mass of non-BCG galaxies within a given 3D aperture (we use primarily $R_{500}$ in this study). The uncertainties in the model parameters ($A$, $R_s$, $\gamma$, and $B$) are fully propagated into the uncertainties in the 3D stellar mass contained in the stellar masses of non-central satellite galaxies.

\subsection{From luminosities to stellar mass}

Converting luminosity to stellar mass requires assumptions about stellar mass-to-light ratio based either on the independent dynamical estimate or on the stellar population synthesis modelling. We note that the current statistical uncertainties for mass-to-light ratios derived by both dynamical and stellar population synthesis (SPS) techniques are substantial  \citep[$\gtrsim 0.1$ dex,][]{bell_dejong01,bell_etal03,cappellari_etal06,zibetti_etal09,conroy13}. The main uncertainty, however, is systematic and is due to uncertainty in our knowledge of the IMF and its variations with galaxy properties.  To facilitate comparison with $M_\ast-M$ relation calibrations from other studies, we will adopt the mass-to-light ratios corresponding to the \citet{chabrier03} IMF as our fiducial choice, but will discuss effects of the IMF variation in \S~\ref{sec:imf}. 
 
\citet{cappellari_etal06} present comparisons of the $I$-band mass-to-light ratios derived from dynamical analysis to those derived from the stellar population synthesis (SPS). The largest values correspond to the most massive galaxies, which have maximum SPS $M/L_I\approx 3$ for the Kroupa IMF or $M/L_I\approx 3/10^{0.05}\approx 2.7$  for the Chabrier IMF \citep[see, e.g., Table 2 in][]{bernardi_etal10}. This is similar to the value of $M_\ast/L_I=2.65$ adopted by
\citet{gonzalez_etal13}, who have derived it using luminosity-dependent dynamical mass-to-light ratio of \citet{cappellari_etal06} and averaged it over $I$-band luminosity function for galaxies with $L_I>0.25L_{\ast,I}$. The average is dominated by $L\approx L_\ast$ galaxies for which dynamical mass-to-light ratios are consistent with those expected for the Chabrier IMF. 

To estimate stellar masses from the luminosities measured from SDSS
data for nine clusters described above, we have used relation between
$M/L$ and galaxy color derived from the stellar population synthesis
(SPS) models of \citet{bell_etal03}: $\log_{10}(M_\ast/L_r)_{\rm pop}
= 1.097\times (g-r)-0.306$ and $\log_{10}(M/L_i)_{\rm pop} =
0.864\times(g-r)-0.222$. Massive cluster galaxies exhibit
substantial radial color gradients. Therefore, we use color profile to
convert luminosities to stellar mass at each particular radius.  The
projected stellar surface density profiles were thus obtained from the
surface brightness profiles, $\Sigma_{M_\ast}=(M/L)_{\rm pop}(R)\times
\Sigma_L(R)$, for $r$ and $i$ bands. The total stellar mass was obtained by integrating the
profiles to infinity using the best fit three-component S\'ersic model for $\Sigma_L(R)$.

\begin{table*}[t]
\begin{center}
\caption{Nine galaxy clusters with the SDSS stellar mass measurements}
\label{tab:sdsscl}
\begin{tabular}{rcrrrlrrrr}
\hline\hline\\
Cluster name & $z$ & $\Rfh$ & $\Mfh$ &  $R_{\rm out}$ & $M_{\ast,\rm BCG}(<R_{\rm out})$ & $M_{\ast,\rm BCG}$  &   $M_{\ast,\rm non-BCG}$ & $r_{1/2}$  & $r_{1/2}(<R_{\rm out})$ \\ 
         &    &  kpc  &   $10^{14}\ \rm M_{\odot}$ &  kpc & \multicolumn{3}{c}{$10^{12}\ \rm M_{\odot}$} & \multicolumn{2}{c}{kpc}\\
\\\hline\\
A2142 & 1539 & $11.96  \pm 0.200$ & $15.60$ &  220 &    $1.95 \pm 0.17^a$ &    $3.12 \pm 0.36$ &     $12.22 \pm 1.58$ &  $ 179 \pm   41$ &  $  52 \pm    9$\\
A2029 & 1387 & $8.64  \pm 0.140$ & $10.30$ &  271 &    $3.42 \pm 0.20$ &    $4.14 \pm 0.30$ &     $8.21 \pm 1.05$ &  $  84 \pm    7$ &  $  44 \pm    6$\\
A85 & 1235 & $5.98  \pm 0.070$ & $7.00$ &  308 &    $2.49 \pm 0.21$ &    $3.06 \pm 0.30$ &     $5.28 \pm 0.74$ &  $ 174 \pm   24$ &  $  69 \pm   19$\\
A1795 & 1196 & $5.46  \pm 0.060$ & $5.34$ &  210 &    $1.22 \pm 0.09$ &    $1.47 \pm 0.13$ &     $4.01 \pm 0.75$ &  $  99 \pm   10$ &  $  47 \pm    9$\\
MKW3s &  873 & $2.09  \pm 0.030$ & $2.35$ &  116 &    $0.73 \pm 0.03^b$ &    $0.79 \pm 0.05$ &     $1.89 \pm 0.49$ &  $  34 \pm    4$ &  $  22 \pm    3$\\
A2052 &  840 & $1.84  \pm 0.030$ & $1.86$ &  186 &    $1.08 \pm 0.07$ &    $1.26 \pm 0.11$ &     $2.22 \pm 0.47$ &  $  53 \pm    3$ &  $  31 \pm    4$\\
A1991 &  748 & $1.33  \pm 0.037$ & $1.34$ &  117 &    $1.05 \pm 0.05$ &    $1.09 \pm 0.06$ &     $1.77 \pm 0.41$ &  $  37 \pm    2$ &  $  25 \pm    2$\\
MKW4 &  568 & $0.56  \pm 0.013$ & $0.46$ &  133 &    $0.87 \pm 0.04$ &    $0.91 \pm 0.05$ &     $0.97 \pm 0.29$ &  $  29 \pm    2$ &  $  20 \pm    2$\\
RXJ1159 &  568 & $0.59  \pm 0.028$ & $0.47$ &  142 &    $1.10 \pm 0.05$ &    $1.38 \pm 0.14$ &     $0.47 \pm 0.18$ &  $  31 \pm    5$ &  $  15 \pm    2$\\
  \\
\hline
\end{tabular}
\end{center}
\tablecomments{$^a$ The stellar masses of the second brightest galaxy in A2142 are: $M_{\ast}(<R_{\rm out})=7.8\pm 0.43\times 10^{11}\ \rm M_{\odot}$, where $R_{\rm out}=101$ kpc, and total mass extrapolated to infinity $M_{\ast,\rm tot}=8.50\pm 0.63\times 10^{11}\ \rm M_{\odot}$; $^b$ the stellar masses of the second brightest galaxy in MKW3s are: $M_{\ast}(<R_{\rm out})=3.6\pm 0.16\times 10^{11}\ \rm M_{\odot}$, where $R_{\rm out}=68$ kpc, and total mass extrapolated to infinity $M_{\ast,\rm tot}=3.78\pm 0.39\times 10^{11}\ \rm M_{\odot}$.}
\end{table*}

The mass-to-light ratios derived by \citet{bell_etal03}
correspond to the ``diet Salpeter'' IMF and are $\approx 0.1$ dex
above the mass-to-light ratios expected for the Chabrier IMF
\citep{bell_dejong01,bell_etal03,bernardi_etal10}. We therefore adjust
the resulting stellar masses down by 0.1 dex. We have checked that we
get very similar stellar masses for the brightest cluster galaxies if
we use a constant value $M_\ast/L_r=3.0$ which is predicted for the
Chabrier IMF for the typical color $g-r\approx 0.8$ of massive
ellipticals in our sample \citep[see the $M/L$-color relations in the
Appendix of ][]{zibetti_etal09}. We have also checked that we obtain
indistinguishable stellar masses from $r$- and $i$-band
luminosities. We will present $r$-band derived stellar masses in the
remainder of this paper.

Overall, there is a reasonably good agreement
between our derived BCG stellar masses and those derived by G13, 
although stellar masses measured by G13
appear to be somewhat higher for low mass clusters,
$\Mfh\approx 10^{14}\ \Msun$. It is not clear whether this difference is systematic given a small
number of objects in the two samples. 
The background level is comparable in both analyses and thus difference in 
surface density sensitivity unlikely plays a role. Nevertheless, 
the outer profiles of BCGs in the G13 sample are typically traced to radii $\approx 300$~kpc, compared to a
typical outer radii of $\sim 100-200$~ kpc for our
clusters.\footnote{Only in two clusters, Abell 2029 and 85, we trace
  profile significantly beyond $200$ kpc, out to $270$ and $340$ kpc,
  respectively.} This difference can be due to a higher amplitude of the outer surface density components in their systems, as this amplitude varies from cluster to cluster. 
  Significant contribution to the total stellar mass of BCGs comes from extrapolation of this
  outer component to infinity and thus even modest differences in amplitude 
for some clusters can lead to some difference in the derived total stellar mass. 
We have compared the masses measured in the two studies within a fixed aperture of 50 kpc, not affected by outer regions and profile extrapolations. The masses
measured within this aperture are in good agreement, which implies that there are no systematic differences due to different mass-to-light ratios. In the reminder of this paper we will use both G13 and our data jointly in most of the analyses.

\section{Stellar content of high mass halos}
\label{sec:msmh}
 In this
section we explore the connection between bulk properties of massive halos, 
such as their mass and radius, and stellar masses and sizes of galaxies they host. 
We will also consider how these properties of high mass halos are related
to the corresponding properties of smaller, galaxy-scale halos. Before we can consider such relation, we need to identify appropriate counterparts to the central galaxies in galaxy-sized halos among cluster galaxies. 

Low and high mass halos differ qualitatively in the way stellar mass is distributed.  In galaxy-sized halos most of the stellar mass is concentrated in the central galaxy, while in cluster-sized halos significant fraction of stars is 
in satellite galaxies. Nevertheless, brightest cluster galaxies occupy a special place among galaxies 
in cluster-sized halos. First, in most X-ray clusters BCGs are located close to the center of cluster halo \citep[e.g., $\approx 80\%$ of BCGs are found within $0.1R_{200}$ of the peak of X-ray emission,][]{lin_mohr04}. 
They could thus be most naturally identified as analogs of central galaxies in smaller mass halos \citep[although note that not all of the BCGs are closest to the center, e.g.,][]{skibba_etal11}. Second, as we discussed in Section~\ref{sec:data} above, distribution of stars associated with BCGs extends to hundreds 
of kpc -- i.e., a significant fraction of the cluster virial radius. The outer regions of BCGs are often called the intracluster light  and are interpreted as a separate component forming via mergers and tidal disruption of satellite galaxies. 

Although the stars in the outskirts of the BCGs may indeed have a different origin than stars in the inner regions,  Figure~\ref{fig:bcgpror2} shows that surface density profiles of BCGs  extend smoothly to the largest reliably probed radii. Although profiles do have weak features, which require multiple components for accurate modelling, these features are subtle and occur at different radii in different objects. The profiles do not show clearly distinct components with very different profiles, which can justify clean separation similar, for example, to the bulge-disk decomposition in late type galaxies. Thus, in practice, the outer component of BCGs cannot be cleanly and reliably separated based on the surface density profiles alone. Such decomposition would also create questions about comparisons of stellar masses derived for BCGs and stellar masses of smaller mass galaxies, for which no such separation is done when their luminosity and stellar mass are estimated. For example, \citet{bernardi_etal13} fit single S\'ersic component to all SDSS galaxies and derive stellar mass function that does not contain any features which would be associated with emergence of a distinct ICL component unique only to cluster halos. As we have shown above in Section~\ref{sec:lcomp} (see Fig.~\ref{fig:bcgcompl}), such single S\'ersic component fits recover luminosity of BCGs similar on average to that recovered using a more detailed with with three S\'ersic components. Finally, removal of the outer component creates problems for comparisons
with  theoretical models, as this would require detailed modelling of the BCG profiles to closely mimic the observational component decomposition procedure. This is often difficult or impossible in practice. 

Given these considerations, we will not attempt to separate
the outer component of BCGs in our analysis below.  We also advocate defining stellar mass in well specified apertures fixed in physical scale or at a fixed fraction of virial radius and provide such mass measurements for our BCGs in the Appendix~\ref{sec:altmass}.

\subsection{BCG stellar mass--halo mass relation}
\label{sec:mscmh}

\begin{figure}[t]
\vspace{-1cm}
\begin{center}
\includegraphics[scale=0.475]{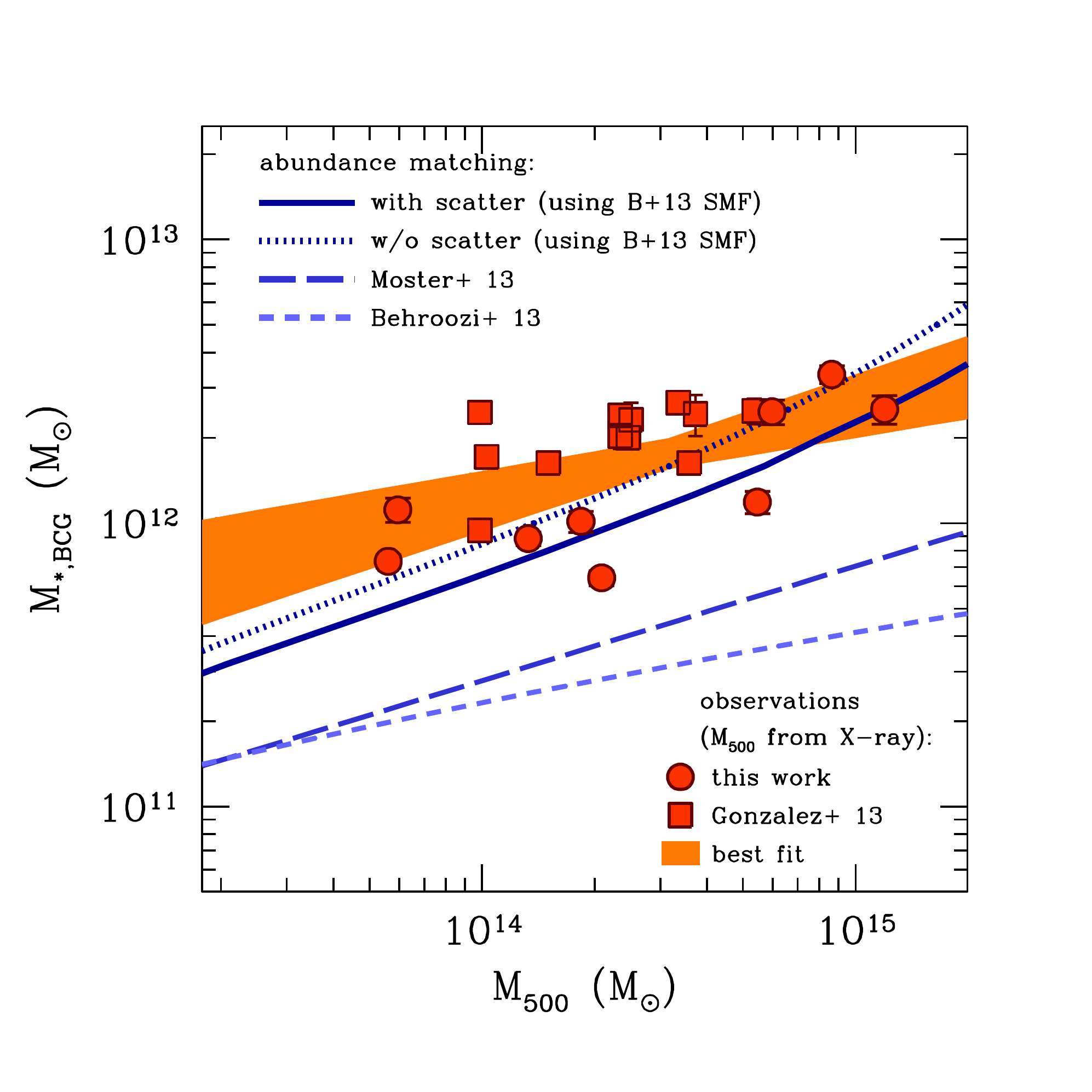}
\vspace{-0.8cm}
\caption{Stellar masses of BCG galaxies versus the total halo mass, $\Mfh$. Circles show the BCGs in the clusters analyzed in this paper, while squares show BCGs from the recent analysis of \protect\citet{gonzalez_etal13}. The orange band shows the best fit power law relation and its $1\sigma$ uncertainties in normalization and slope (see Table~\ref{tab:bestfit}). The solid and dotted lines show relation derived using the abundance matching approach with and without scatter (i.e., assuming a scatter of 0.2 dex and zero in stellar mass at fixed halo mass) using stellar mass function measured by \protect\citet{bernardi_etal13}. The long and short dashed curves show relations derived by \protect\citet{moster_etal13} and \protect\citet{behroozi_etal13a} using abundance matching approach, but using different stellar mass functions. The total halo masses presented in the relations from the latter two studies were converted to $\Mfh$ using average halo mass-concentration relation.} 
\label{fig:msmhhigh}
\end{center}
\end{figure}

Figure~\ref{fig:msmhhigh} shows relation between stellar mass of the
brightest cluster galaxies and cluster mass, $\Mfh$, for the nine
clusters analyzed in this study and twelve clusters from the study of
\citet{gonzalez_etal13}. Although the scatter in $M_{\ast}$ at a fixed $\Mfh$ is significant, there is an overall trend for BCG stellar mass  to increase with total cluster mass. A power law fit to the trend is: $\log_{10} \Mscen=12.25\pm 0.044 + (0.34\pm 0.11)(\log_{10}M_{500}-14.5)$, where masses are in $M_{\odot}$. In this and other fits below, we have derived the slope, normalization, and scatter using likelihood maximization   taking into account both errors in $\Mfh$ and $M_{\ast,\rm BCG}$ \citep{hogg_etal10}. The procedure fits for the scatter in the direction perpendicular to the power law line. However, we report scatter in $M_\ast$ at a fixed $M$, which we derive by projecting scatter onto the vertical direction using 
the final best fit slope. The results of this and other fits are summarized in Table~\ref{tab:bestfit}. The best fit slope of the relation is similar to the slope $0.33\pm 0.06$ of the relation of $K$-band BCG luminosity and $M_{200}$ found by \citet{lin_mohr04} for clusters in the same mass range. Our measured scatter of $\approx 0.17$ dex is somewhat larger than the $\approx 34\%$ fractional scatter found by these authors, but the difference is not very significant compared to uncertainty of scatter. Note that the scatter we measure at the cluster mass scale is comparable to the scatter estimated at the galaxy mass scale by \citet[][]{more_etal11a} in their analysis of satellite kinematics.
 
In addition to observational points, figure~\ref{fig:msmhhigh} shows the $M_*-M_{500}$ relations expected from the abundance matching ansatz,  in which relation between total halo mass, $M$, and stellar mass of galaxies they host, $M_*$, is established implicitly by matching cumulative stellar and halo mass functions.   
In the simplest implementation of this model, the cumulative abundances are matched directly, assuming no scatter in the relation between stellar and halo masses: $n(>M_*)=n(>M)$. However, it is straightforward to include scatter in this approach (see Appendix A). Figure~\ref{fig:msmhhigh} shows that relations predicted by the AM ansatz in recent studies\footnote{These studies present relation for the halo masses defined using overdensity of $200\rho_{\rm crit}$ and virial overdensity, respectively. We have converted these masses to the overdensity of $500\rho_{\rm crit}$ assuming NFW profile form and $c-M_{\rm vir}$ relation predicted by the model of \citet{bullock_etal01} but a constant minimum ``floor'' of concentration $c_{\rm vir}=3.5$ at large masses \citep{zhao_etal03b}.} of \citet[][]{moster_etal13} and \citet{behroozi_etal13a}, shown by the long and short dashed lines, underestimate stellar masses at a given halo mass for such massive galaxies by a factor of up to $\gtrsim 5$. The reason for this discrepancy is significant underestimate of luminosities and stellar masses for high-mass galaxies in the stellar mass functions adopted to derive these relations, as was shown recently by \citet{bernardi_etal13} and as we discussed above in \S~\ref{sec:lcomp}. 

The solid and dotted lines in Figure~\ref{fig:msmhhigh} show results of abundance matching with and without scatter in the $M_\ast-M$ relation, as described in the Appendix, using the \citet{bernardi_etal13} stellar mass function derived from the luminosity functions with corrected photometry for the luminous galaxies. As could be expected from the good agreement between \citet{bernardi_etal13} luminosities and luminosities measured in our study (see Figure~\ref{fig:bcgcompl}), these abundance matching results are in considerably better agreement with observed  $M_\ast-M$ relation for the brightest cluster galaxies.

\subsection{Radial profiles and sizes of the BCG galaxies}
\label{sec:bcgpro}

Results in the previous section show that stellar masses of BCG galaxies correlate with the total stellar mass of their parent halo with scatter comparable to that for the MW-sized galaxies. Here we examine whether characteristic {\it size} of stellar distribution of BCGs correlates with halo virial radius. 
It is well known that galaxies exhibit a correlation between their half-mass radii and stellar masses \citep[e.g.,][and references therein]{bernardi_etal12}. The relation is
complicated, however, and cannot be described by a single power law across a wide range of masses. 

Given that there is abundant evidence that stellar masses of galaxies correlate with total masses of their host halos, the correlation of sizes with halo mass and virial radius is also expected. What is remarkable, however, is that the resulting relation between galaxy sizes and  virial radii of their host halos is close to {\it linear} over the entire range of stellar masses \citep{kravtsov13}. This 
indicates that sizes of galaxies are set by properties of baryons tightly connected to properties of dark matter,  while the complicated size--stellar mass relation is simply a reflection of
the complicated non-linear $M_{\ast}-M_{200}$ relation resulting from a complicated interplay between star 
formation and feedback during galaxy evolution. 
Here we examine how sizes of BCGs, as galaxies representing the most extreme massive 
end of galaxies, fit into this picture. 

\begin{figure}[t]
\begin{center}
\hspace{-0.5cm}\includegraphics[scale=0.475]{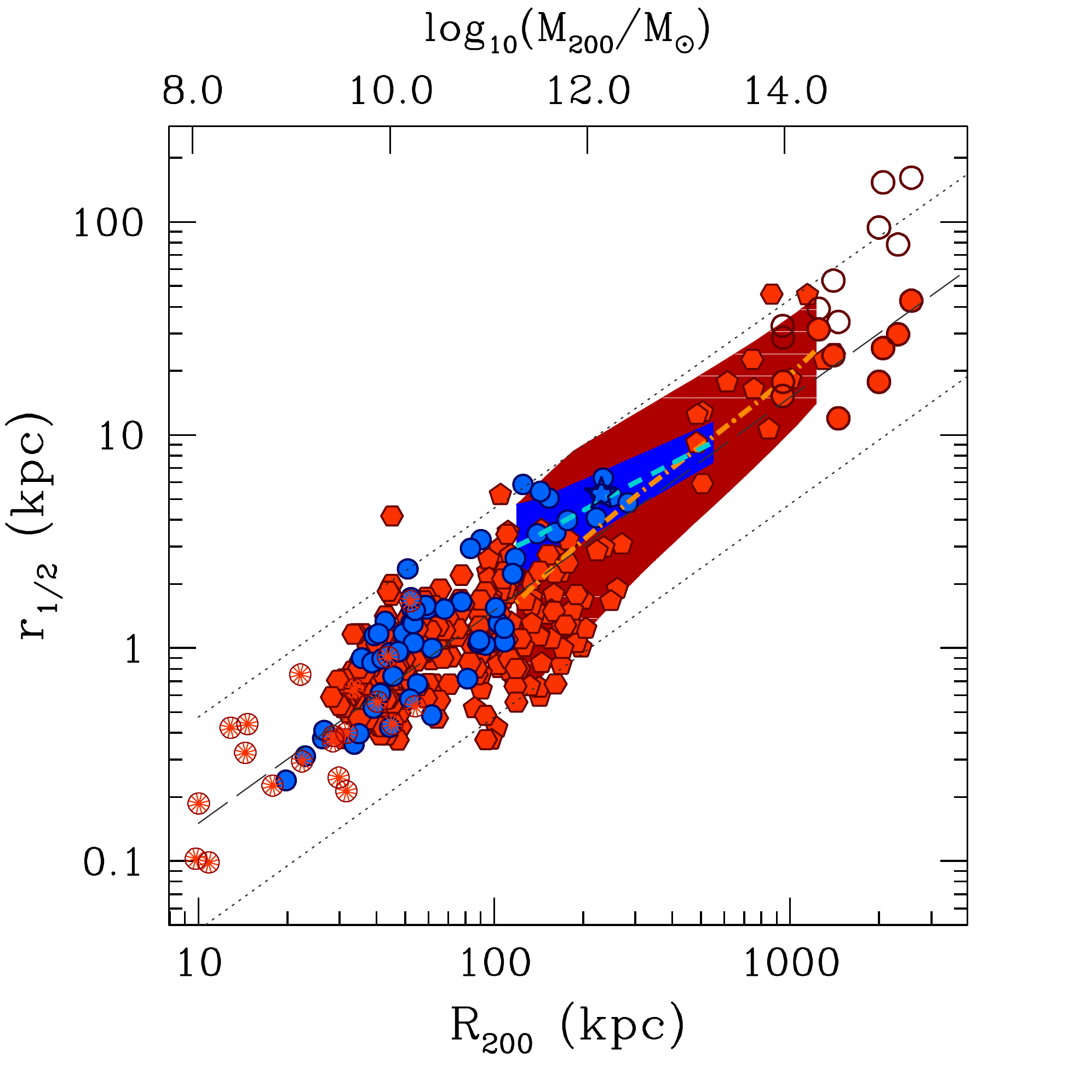}
\caption{Relation between three dimensional half-mass radii of stellar distribution in galaxies  and virial radii of their parent halos, $\Rtwoh$, defined as the radius enclosing overdensity of $200\rhoc$. The {\it open red circles} show the nine BCGs analyzed in this paper, with $r_{1/2}$ estimated using the entire stellar surface density profile, as described in the text. The {\it solid red circles} show the corresponding radii only for the combined inner and middle best fit S\'ersic components. The other {\it red points }  show samples of elliptical, dwarf elliptical, and dwarf spheroidal galaxies from the compilation of \protect\citet{misgeld_hilker11}; {\it blue circles} are the late type galaxies from the samples of \protect\citet{leroy_etal08} and \citet{zhang_etal12}. The {\it light blue dashed line} and {\it dot-dashed orange line} show the average relations derived for late and early-type galaxies, respectively, from the average $R_{\rm 1/2}-M_*$ relations of \protect\citet{bernardi_etal12}, while {\it blue and dark red shaded bands} show $2\sigma$ scatter around these mean relations. {\it The gray dashed line} shows linear relation $r_{1/2}=0.015\Rtwoh$ and dotted lines are linear relations offset by 0.5 dex, which approximately corresponds to the scatter in galaxy sizes from distribution of  halo spin parameter $\lambda$ under assumption that $r_{1/2}\propto\lambda\Rtwoh$.} 
\label{fig:rhrg}
\end{center}
\end{figure}

\begin{figure}[t]
\begin{center}
\vspace{-0.5cm}
\includegraphics[scale=0.485]{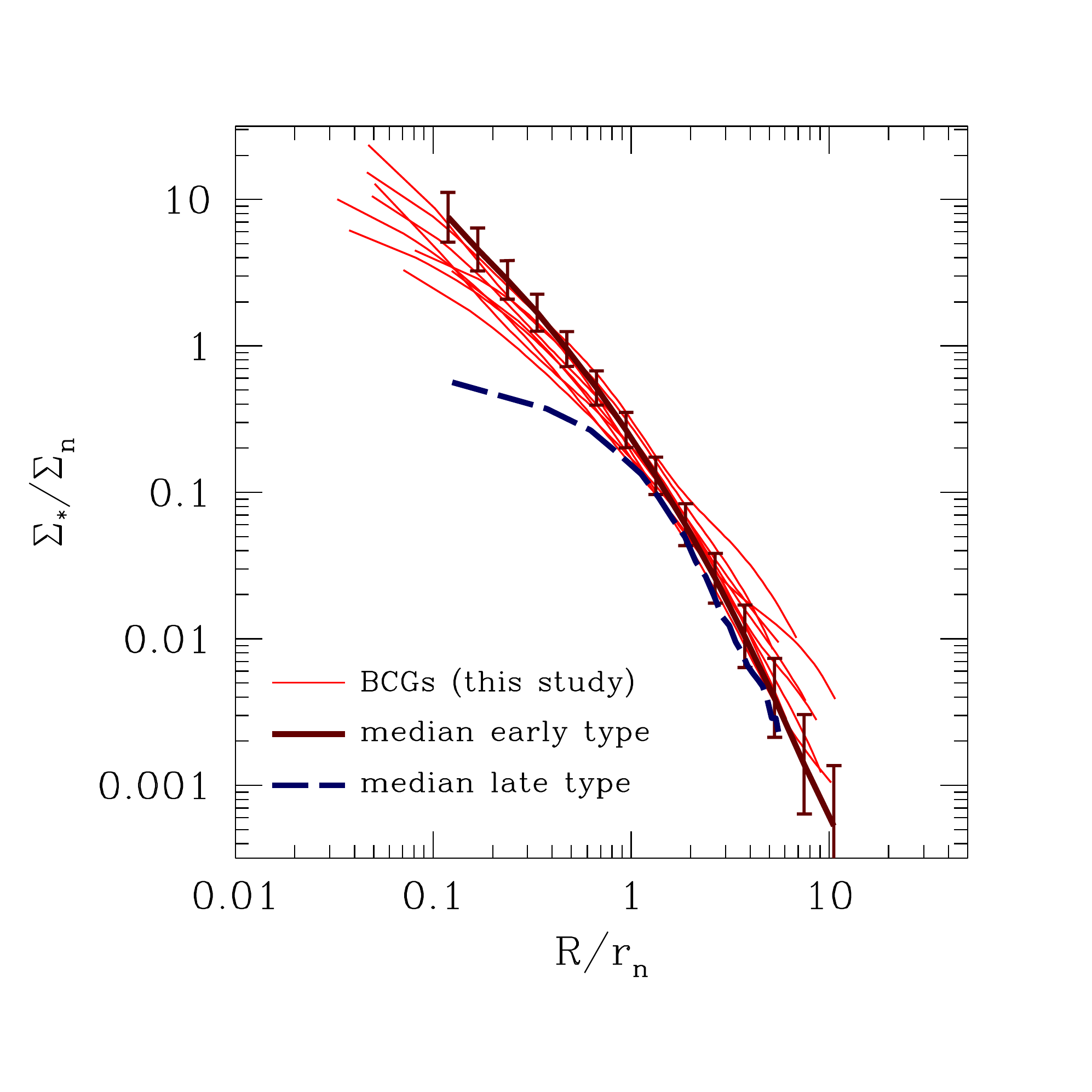}
\vspace{-1cm}
\caption{Normalized surface mass density profiles of nine BCGs in our sample (thin red lines) compared to the median profile of massive early type (S\'ersic index $n>2.5$) SDSS galaxies from \protect\citet{szomoru_etal13} and late type galaxies from the THINGS and LITTLE THINGS samples \protect\citep[blue, dashed line, see][for details]{kravtsov13}.  Profiles are normalized by the radius $r_n=0.015R_{200}$ which approximately corresponds to the half mass radius of stellar distribution. Here for smaller galaxies $R_{200}$ is obtained using the abundance matching ansatz from galaxy's $M_*$, while for BCGs we use $R_{200}=1.67R_{500}$, where $R_{500}$ is measured from X-ray data. The surface densities of galaxies are normalized by $\Sigma_{\rm n}=M_{\ast}/(\pi R_{\rm eff})^2$, where $R_{\rm eff}$ is the half mass radius of stellar distribution. For the BCG galaxies $M_{\ast}$ and $R_{\rm eff}$ are the stellar mass and half mass radius of the inner and middle S\'ersic components (see \S~\ref{sec:bcgpro} for details).  The error bars around thick red lines show rms dispersion around the median. The figure shows that the inner regions of the BCGs have profiles that are scaled versions of the profiles of smaller mass early type galaxies. Remarkably, at $R\gtrsim r_{\rm n}$ profiles of all late and early type galaxies have similar shape.} 
\label{fig:spronorm}
\end{center}
\end{figure}

Figure~\ref{fig:rhrg} shows the relation between three dimensional half-mass radius of stellar distribution and virial radius of their halo, $R_{200}$. The latter is estimated using $M_{\ast}-M_{200}$ relation derived from abundance matching and $R_{200}$ calculated from $M_{200}$ for all galaxies, except for BCGs for which we use $R_{200}\approx 1.67R_{500}$ appropriate for NFW profile with a typical concentration for cluster halos. The abundance matching is carried out using the \citet{bernardi_etal13} SMF with 0.2 dex scatter in $M_\ast$ at a fixed $M_{200}$, as described above (see also Appendix \ref{sec:amapp}). The sample of late type and early type galaxies shown in the figure is identical to that presented in \citet{kravtsov13}. Here we also add half-mass radii of the BCGs in the nine clusters analyzed in this paper. To estimate the half mass radii we use the three-component S\'ersic fit to the stellar surface density profiles and measure the projected radius containing half of the total stellar mass measured by extrapolating the best fit three-component fit to infinity. 
 The projected effective radius is then converted into the 3D half-mass radius using $\rhalf=1.34\Reff$, the expression accurate for spheroidal systems described by the S\'ersic profile with a wide range of the S\'ersic index values \citep[see eq. 21 in][]{limaneto_etal99}.  
The 3D half-mass radii measured in this way are shown in Figure~\ref{fig:rhrg} as a function of  $R_{200}$  as open circles. 

We can see that the half-mass radii of BCGs lie above the average $r_{1/2}\approx 0.015R_{200}$ relation but are still within the scatter exhibited by smaller mass galaxies and scatter expected from the distribution of spins if sizes are set by specific angular momentum shown by the dotted lines. However, one should note that the actual half mass radii could be larger if the extrapolation of the triple-S\'ersic profile underestimates stellar mass profile beyond the 
largest radii of our measurements. As we discuss in Section \ref{sec:sdssbcgmass}, such underestimate should have an effect of $\lesssim 30\%$ on the total BCG stellar mass.
The half-mass radius can, nevertheless, be significantly affected. One should therefore view the open circles in Figure \ref{fig:rhrg} as lower limits. 

Recently, \citet{huang_etal13} have used three-component S\'ersic fits for a sample of nearby massive elliptical galaxies and showed that the relation between half mass radii and stellar masses of the inner two components resembles that size-stellar mass relation of $z\approx 1$ compact red galaxies. They proposed an evolutionary scenario in which the inner regions of massive ellipticals form by dissipative collapse of baryons at high $z$, while outer regions form by accretion and mergers at later epochs. We show the half mass radii and stellar masses of the inner plus middle S\'ersic components for the nine BCGs analyzed in this study by solid red circles in Figure~\ref{fig:rhrg}. As we noted above, separation into separate components based on the 
surface density profiles is likely to be noisy. Nevertheless, the figure shows that the sizes and stellar masses of the combined two inner S\'ersic components are scattered around the average $r_{1/2}\approx 0.015R_{200}$ relation and are thus consistent with the size--virial radius correlation of smaller mass galaxies. This means that the inner regions of BCGs could indeed form via processes similar to those shaping structure of smaller mass
galaxies. This can be seen even more clearly if we compare stellar surface density profiles of 
BCGs with those of smaller mass galaxies.  

Figure~\ref{fig:spronorm} shows the normalized surface mass density profiles of nine BCGs in our sample compared to the median profile of massive early type (S\'ersic index $n>2.5$) SDSS galaxies from \citet{szomoru_etal13} and late type galaxies from the THINGS and LITTLE THINGS samples derived in \citet[][]{kravtsov13}. The radii are normalized by the radius $r_n=0.015R_{200}$, while the surface densities of galaxies are normalized by $\Sigma_{\rm n}=M_{\ast}/(\pi R_{\rm eff})^2$, where $R_{\rm eff}$ is the half mass radius of stellar distribution. For the BCG galaxies, $M_{\ast}$ and $R_{\rm eff}$ are the stellar mass and half mass radius of the inner and middle S\'ersic components.  The error bars around thick red lines show rms dispersion around the median. The figure shows that the inner regions of the BCGs have profiles that are scaled versions of the profiles of smaller mass early type galaxies. Remarkably, at $R\gtrsim r_{\rm n}$ profiles of all early {\it and late} type galaxies have similar shape \citep[see also][]{kravtsov13,fang_etal13}.

Results presented in these figures show a striking similarity between the inner regions of all elliptical galaxies. Remarkably, the sizes of late type, early type and the inner regions of BCG galaxies all corresponds to a similar fraction of the virial radius, which indicates that sizes of all galaxies may be set by a similar mechanism. Indeed, as discussed in \citet{kravtsov13}, the average $r_{1/2}\approx 0.015R_{200}$ is consistent with expectations of the \citet{mo_etal98} model in the case when the specific angular momentum of baryons is similar to that of dark 
matter. The BCGs do have the outer extended envelope that likely forms by mergers and tidal stripping of satellite galaxies \citep[e.g.,][]{murante_etal07,conroy_etal07,purcell_etal07,puchwein_etal10,feldmann_etal10,watson_etal12} and can thus be significantly larger than the size defined by the angular momentum limit. Our results
thus are consistent with scenario, in which the inner stellar distribution of BCGs is set early via dissipative processes, while the outer regions are built up over longer period of time by mergers and accretion. The sizes of BCGs are correlated with the virial radii of their parent cluster 
and lie on the extension of the corresponding relation for smaller mass galaxies. 
 
\begin{figure}[t]
\vspace{-1cm}
\begin{center}
\includegraphics[scale=0.475]{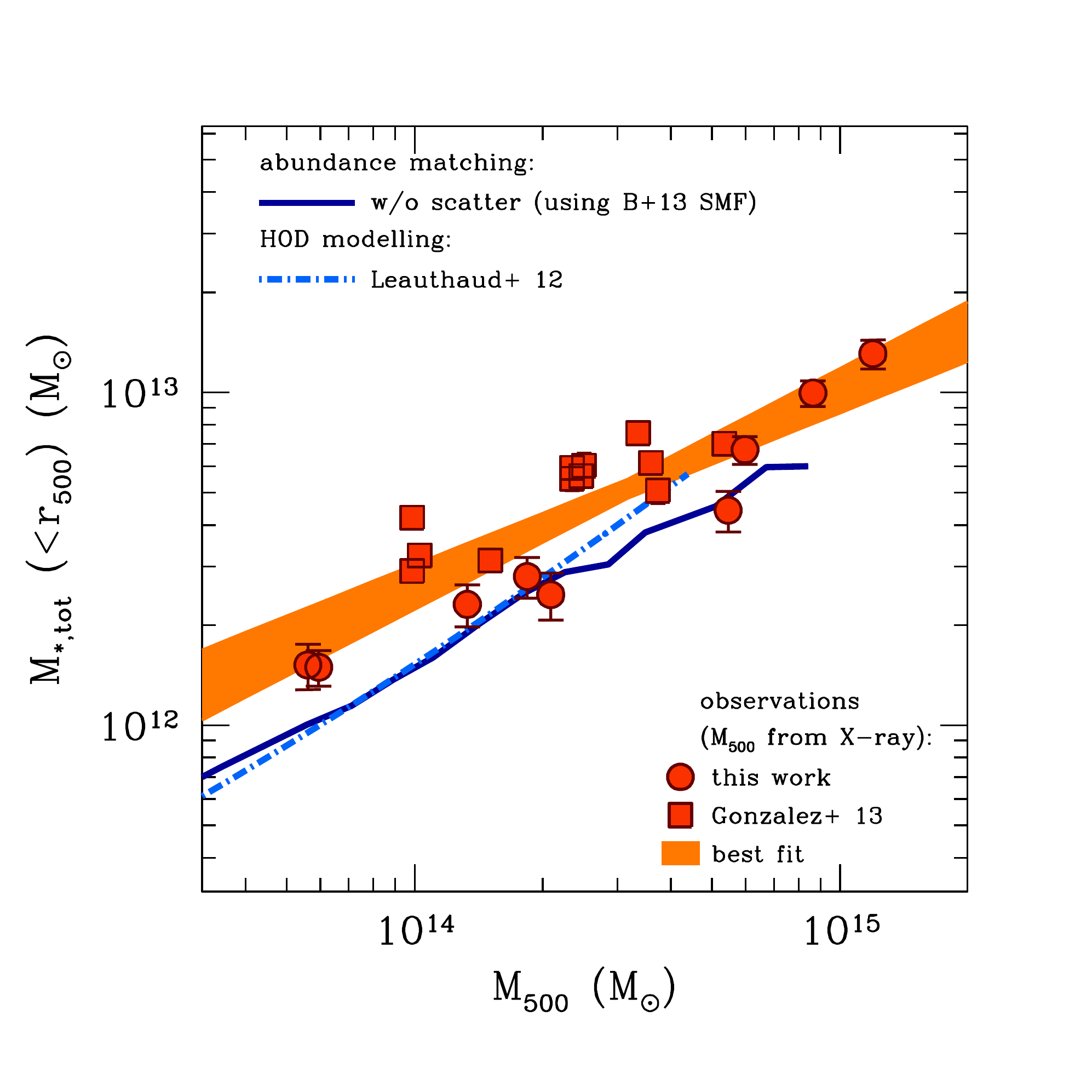}
\vspace{-0.8cm}
\caption{Total stellar mass within $\Rfh$ versus total halo mass, $\Mfh$. Circles show the clusters analyzed in this paper, while squares show clusters from the recent analysis of \protect\citet{gonzalez_etal13}. The orange band shows best fit power law relation; its width reflects $1\sigma$ uncertainties in normalization and slope. The dot-dashed line shows results of the HOD modelling of \citet{leauthaud_etal12b}. Blue solid line shows relation calculated using halos and subhalos from the Bolshoi simulations with stellar masses assigned using the mean relation derived using abundance matching without scatter. } 
\label{fig:msmhtothigh}
\end{center}
\end{figure}

\subsection{Total stellar mass--halo mass relation}
\label{sec:msmhtothigh}

Having examined correlation of properties of the BCG with cluster mass and virial radius, in this section we investigate correlation of the total stellar mass of clusters with their total mass.
Figure~\ref{fig:msmhtothigh} shows the total stellar mass within $\Rfh$ as a function of $\Mfh$. It is clear that this relation is steeper and tighter than relation for stellar mass of the BCG. The best fit slope is $0.59\pm 0.08$, while the scatter in total stellar mass at a fixed $\Mfh$ is only $0.11\pm 0.02$ (see Table~\ref{tab:bestfit}). The tightness of this relation has been noted by \citet{andreon12}, who quoted 90\% upper limit on scatter of 0.06 dex and argued that total stellar mass can be a good total mass proxy rivaling X-ray proxies, such as gas mass and $Y_{\rm X}$. Using a larger sample, we find a larger scatter of $\approx 29\%$, comparable to scatter of total X-ray luminosity (without core excision) or richness \citep[e.g.,][]{rykoff_etal12} at a fixed mass, but several times larger than scatter in gas mass and $Y_{\rm X}$ \citep[e.g.,][]{kravtsov_etal06,nagai_etal07b}. Our best fit slope is also significantly steeper than the best fit slope of $0.37\pm 0.07$ found by \citet[][]{andreon12}.

Our results are closer to those of \citet{lin_etal04}, who found that the total $K$-band luminosity--$M_{500}$ relation has slope of $0.69\pm 0.04$ and scatter of $\approx 32\%$. 
 Our fit is also close to the best fit relation derived by \citet{lin_etal12} for a  sample of 94 clusters at $0\lesssim z\lesssim 0.6$  for which stellar masses were estimated using IR data from 2MASS survey and the WISE satellite, while total masses were estimated using X-ray mass proxies. In fact, the best fit relation for this sample is a good fit to the nine SDSS clusters analyzed in this study, although normalization is somewhat lower than the best fit for the combined sample of 21 clusters. \citet{lin_etal12} found that normalization of the relation does not evolve with redshift within uncertainties \citep[see also][]{vanderburg_etal13} and thus comparison of their results at higher $z$ to our results at $z<0.1$ is justified. The scatter of $\Mstot$ at a fixed $\Mfh$ in their sample is $0.12\pm 0.01$ in good agreement with our results. 
The $\Mstot-\Mfh$ relation derived by \citet[][shown by the dot-dashed line in Fig.~\ref{fig:msmhtothigh}]{leauthaud_etal12b} based on the HOD analysis of the COSMOS data at $z\approx 0.37$ is somewhat steeper than the best fit of \citet{lin_etal12} and our best fit relation.

\begin{figure}[t]
\vspace{-1cm}
\begin{center}
\includegraphics[scale=0.475]{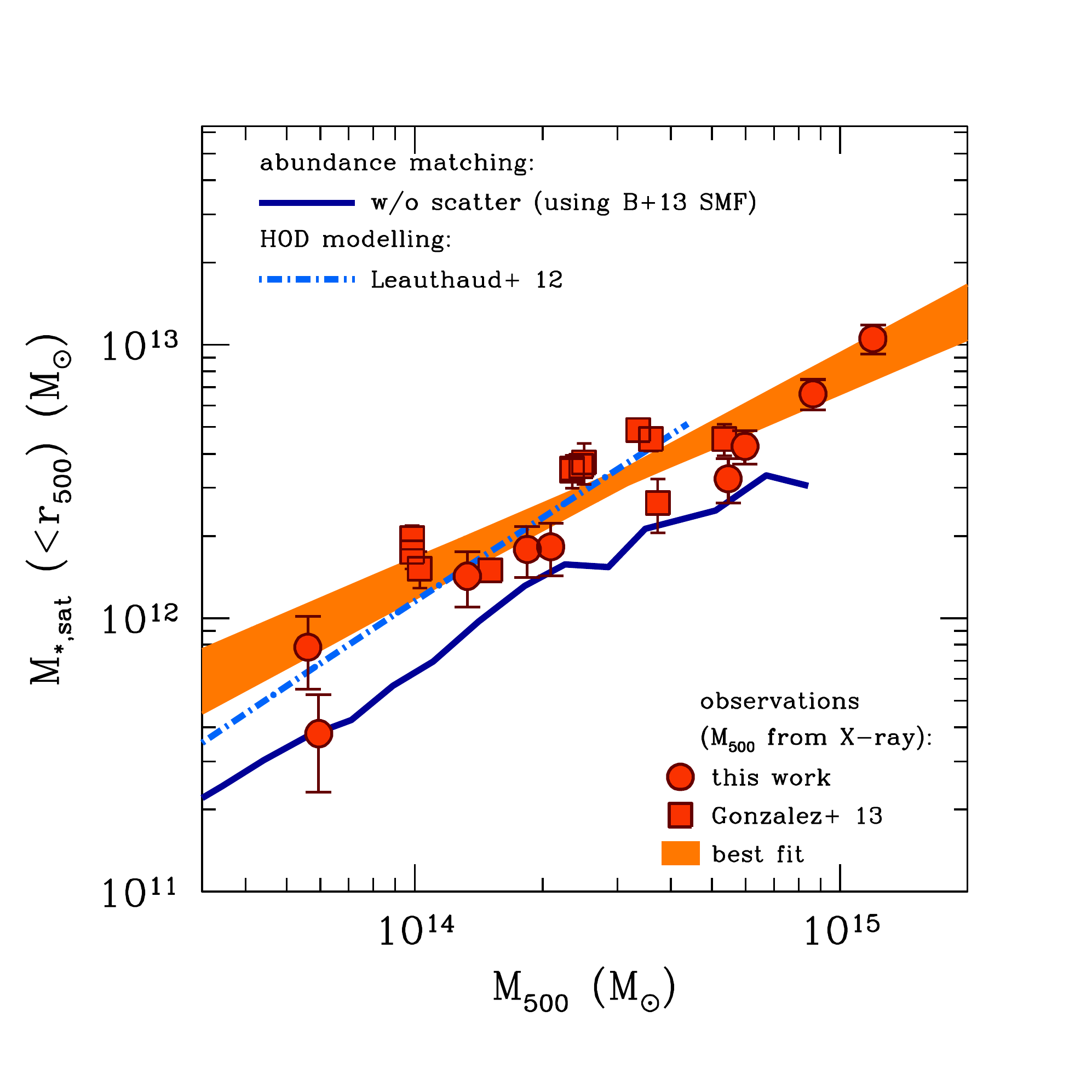}
\vspace{-0.8cm}
\caption{As in Figure~\ref{fig:msmhtothigh}, but for the stellar mass in satellite galaxies within $\Rfh$. } 
\label{fig:msmhsathigh}
\end{center}
\end{figure}

\begin{table}
\begin{center}
\caption{Best fit parameters for power law fits}
\label{tab:bestfit}
\begin{tabular}{cccc}
\hline\hline\\
  Relation  &  slope &   normalization  &  scatter    \\\\\hline\\
             & \multicolumn{2}{c}{9 clusters (this work)} & \\
$\Mscen-M_{500}$  & $0.39\pm 0.17$ & $12.15\pm 0.08$  & $0.21\pm 0.09$  \\
$M_{\ast,\rm sat}-M_{500}$  & $0.87\pm 0.15$ & $12.42\pm 0.07$  & $0.10\pm 0.12$  \\
$M_{\ast,\rm tot}-M_{500}$  & $0.69\pm 0.09$ & $12.63\pm 0.04$  & $0.09\pm 0.05$  \\
\\             
             & \multicolumn{2}{c}{21 clusters (this work $+$ G13)} & \\
$\Mscen-M_{500}$  & $0.33\pm 0.11$ & $12.24\pm 0.04$  & $0.17\pm 0.03$  \\
$M_{\ast,\rm sat}-M_{500}$  & $0.75\pm 0.085$ & $12.52\pm 0.03$  & $0.10\pm 0.03$  \\
$M_{\ast,\rm tot}-M_{500}$  & $0.59\pm 0.08$ & $12.71\pm 0.03$  & $0.11\pm 0.03$  \\
\\
\tableline
\end{tabular}
\tablecomments{The relations are fit by the power law $y=mx+c$, where $x=\log_{10}M_{500}-14.5$   and $y$ is $\log_{10}M_{\ast}$. We have derived these parameters by using likelihood maximization power law fit simultaneously for the slope, normalization, and scatter of the relation taking into account both errors in $\Mfh$ and $M_{\ast}$ \citep{hogg_etal10}. All errors correspond to one standard deviation. The quoted scatter is in $y$ direction in dex. All masses are in $M_{\odot}$ and assume Chabrier IMF.}
\end{center}
\end{table}

 The solid line in Figure~\ref{fig:msmhtothigh}  shows relation calculated using halos and subhalos from the Bolshoi simulation of WMAP compatible cosmology \citep{klypin_etal11} with stellar masses assigned using the mean relation derived from the abundance matching without scatter. Note that for $M_{\ast}-M$ relation of central halo galaxies we use an analytic abundance matching model described in the Appendix~\ref{sec:amapp}, based on the observed SMF, theoretical halo mass function, and parametrization of the fraction of subhalos as a function of halo mass. However, for the total stellar masses of halos, we need to predict subhalo mass functions in host halos of different mass. This can be done either via a halo model or using halo$+$subhalo sample extracted from an actual simulation. Here we choose the latter approach and use the halo catalogs of \citet{behroozi_etal13a}.\footnote{Available at {\tt http://hipacc.ucsc.edu/Bolshoi/MergerTrees.html}}

We use the same $M_{\ast}-M$ relation at $z=0$ derived analytically (see Appendix~\ref{sec:amapp}) to assign stellar masses for the central galaxies of both host halos and subhalos (i.e., halos located within $R_{200}$ of a more massive halo). For host halos we use their current $z=0$ total mass, $M=M_{200}$. To assign stellar mass for subhalos we assume that $M$ is the maximum mass they ever had during their evolution, which was derived from their mass accretion history. This choice is motivated by the comparisons of abundance matching predictions with observed galaxy clustering \citep{reddick_etal13}. Having assigned stellar masses in this way, we then compute total stellar masses of the isolated halos as a sum of stellar masses
 of all subhalos within $R_{500}$ of the center, including stellar mass assigned to the main halo itself. The latter corresponds to the stellar mass of the central halo galaxy.
 
\begin{figure}[t]
\vspace{-1cm}
\begin{center}
\includegraphics[scale=0.475]{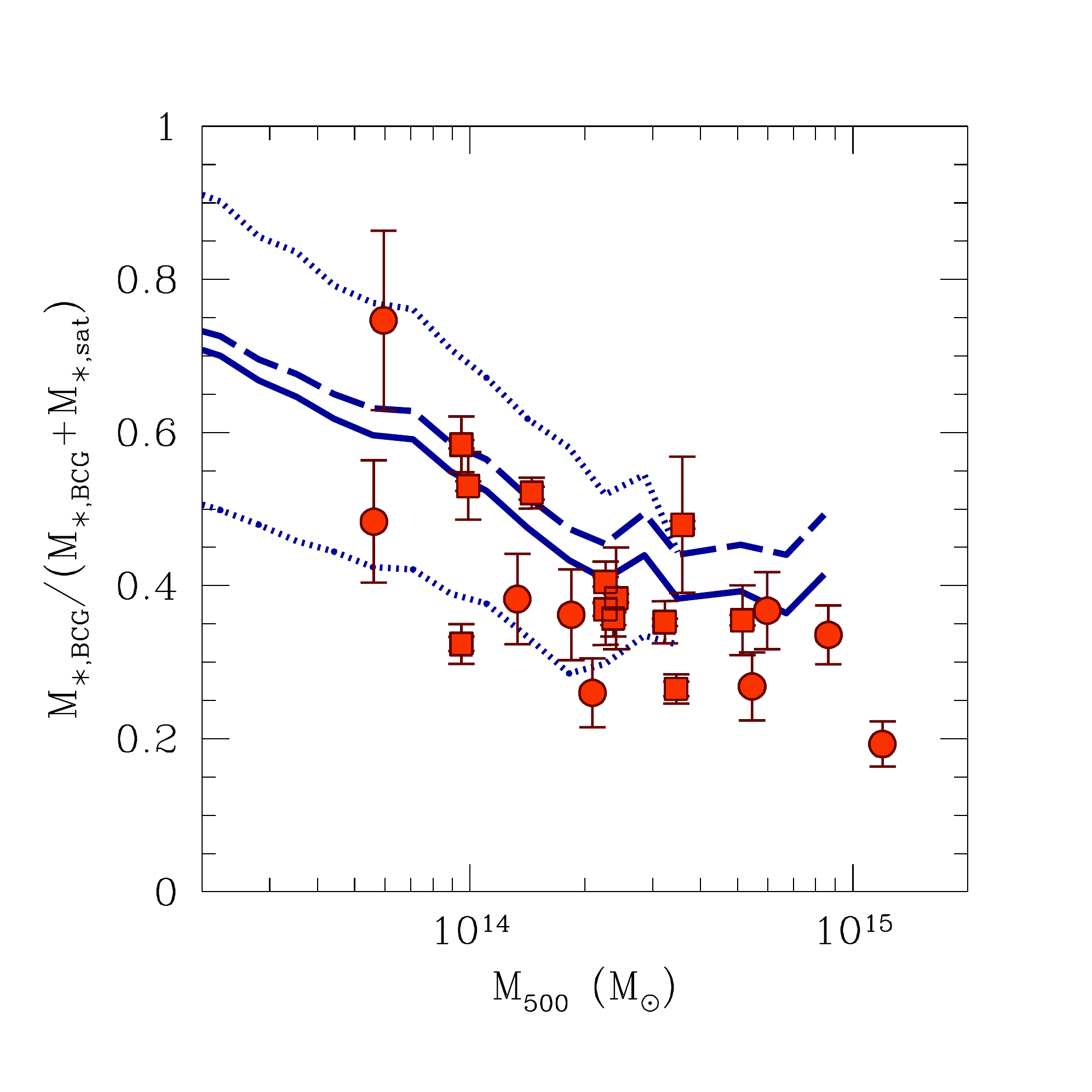}
\vspace{-0.8cm}
\caption{Fraction of the stellar mass in BCG relative to the total stellar mass within $r_{500}$. Circles show the clusters analyzed in this paper, while squares show clusters from the recent analysis of \protect\citet{gonzalez_etal13}. Blue dashed and solid lines show the corresponding fraction derived for halos in the Bolshoi simulation, for which stellar masses were assigned using stellar mass--halo mass relation derived using abundance matching ansatz assuming no scatter and scatter of 0.2 dex, respectively. The dotted lines show rms scatter around the solid line; the last three bins contain only one halo in the Bolshoi catalog and the rms is not defined. } 
\label{fig:fbcg}
\end{center}
\end{figure}
 
 The relation predicted by such AM ansatz has slope similar to our best fit slope for the observed clusters, but normalization lower by $\approx 30-50\%$. The low normalization may be due to several different reasons. The stellar mass--halo mass relation for subhalos may have a higher normalization than normalization for central galaxies. For example, \citet{watson_conroy13} find that for a given mass, subhalos host galaxies with 10\% larger stellar mass compared to host halos of the same mass. Simulation also misses halos and subhalos below resolution limit. For the Bolshoi simulation the overall mass resolution limit for halos is $\approx 2-3\times 10^{10}h^{-1}\,\rm M_{\odot}$, but this limit
 is larger within $\Rfh\approx 0.6R_{200}$ of massive halos due to enhanced tidal disruption in these regions. This incompleteness can account for another $\approx 5-10\%$ of stellar mass. A small difference may also exist between observed masses of clusters and masses assigned via abundance matching, because AM is based on halo mass function calibrated using dissipationless simulations. In simulations with hydrodynamics and galaxy formation $M_{500}$ may be $\sim 5\%$ smaller than in dissipationless simulations \citep{cui_etal12,martizzi_etal13,cusworth_etal14}, which would shift predicted relation left by this amount. These effects can account for a significant fraction of the difference. Despite the discrepancy in normalization, the similarity in slope indicates that overall subhalo mass function and its scaling with host halo mass behave similarly to the real galaxies in clusters. 

Figure~\ref{fig:msmhsathigh} shows relation between stellar mass in satellite galaxies (i.e., the total mass minus the BCG mass) as a function of $\Mfh$. This relation is steeper yet (slope of $0.75\pm 0.09$) and has scatter of $0.10$ dex, comparable to the $\Mstot-\Mfh$ relation. Although $\Mssat-\Mfh$ relation has the slope closest to unity, it is still somewhat shallower than linear scaling. The HOD estimate of \citet{leauthaud_etal12b}, on the other hand, exhibits close to linear scaling \citep[see also][]{skibba_etal07}. Its normalization, however, is close to the normalization of our best fit relation.  

The results presented above show that the total stellar mass in cluster galaxies correlates with total cluster mass with a scatter of $\approx 0.1$ dex, comparable 
to the scatter in richness indicator $\lambda$ \citep{rykoff_etal12}.  
Our results show that the low scatter of the $\Mstot-\Mfh$ relation is due to low scatter of the $\Mssat-\Mfh$ relation, which is shown in Figure~\ref{fig:msmhsathigh}.   Overall, satellite galaxies contribute $\approx 40-50\%$ of the total stellar mass within $\Rfh$ in clusters of mass $\Mfh\approx 10^{14}\rm\ M_{\odot}$, but this fraction increases to $\approx 70-80\%$ in $10^{15}\rm\ M_{\odot}$ clusters. This is shown in Figure~\ref{fig:fbcg}, which also shows abundance matching ansatz reproduces the observed trend reasonably well.

\section{Efficiency of star formation as a function of halo mass}
\label{sec:fstar}

Having quantified the stellar mass--halo mass relation on the cluster mass scale, we will now examine how this relation and the corresponding stellar fraction compare to observational results at smaller masses. Stellar mass fraction quantifies the efficiency with which halos convert their baryons into stars. 
Comparing constraints on cluster and galaxy scales can therefore give us
insights into how star formation efficiency varies from galaxy-sized halos to the largest mass halos in the universe. 

\begin{figure}[t]
\begin{center}
\vspace{-1cm}
\includegraphics[scale=0.475]{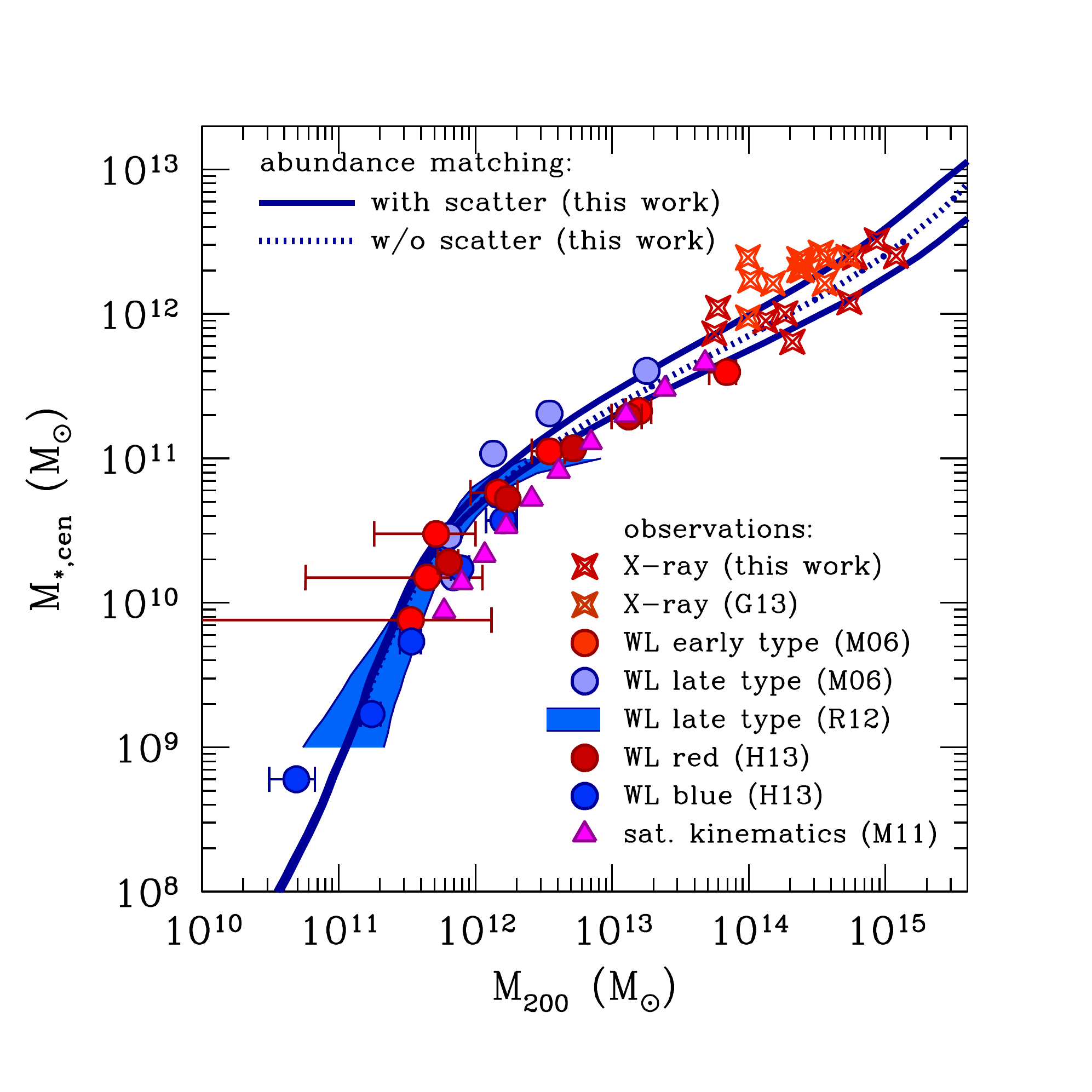}
\vspace{-0.8cm}
\caption{Stellar masses of the central galaxies versus total halo mass, $\Mth$, within the radius enclosing overdensity of 200 times the critical density of the universe. The BCG galaxies in the sample studied in this paper and in the sample of \protect\citet{gonzalez_etal13} are shown by red crosses; cluster masses are converted from $\Mfh$ to $\Mth$ using the concentration--mass relation. Red and blue circles show the mean relations measured using stacked weak lensing signal for early and late type galaxies \protect\citep{mandelbaum_etal06}. Magenta triangles show the mean relation derived using stacked satellite kinematics \protect\citep{more_etal11a}. The solid and dotted lines show relation derived using the abundance matching approach with and without scatter (i.e., assuming a scatter of 0.2 dex and zero in stellar mass at fixed halo mass) using stellar mass function measured by \protect\citet{bernardi_etal13}. Note that this relation is much steeper at the high mass end than previously derived relations using this approach  \protect\citep[][see Appendix~\ref{sec:amapp}]{moster_etal13,behroozi_etal13a} with different stellar mass functions. All stellar masses shown in this figure  were computed assuming Chabrier IMF.}
\label{fig:msmh}
\end{center}
\end{figure}

Figure~\ref{fig:msmh} shows the stellar mass--halo mass relation for a wide range of halo masses and compares results at the high mass end to other observational estimates of this relation at smaller masses, such as weak lensing \citep{mandelbaum_etal06,reyes_etal12,hudson_etal13} and satellite kinematics \citep[][earlier results by \citealt{conroy_etal07b} are consistent with these results and are not shown]{more_etal11a}. As before, we assume that BCGs in clusters correspond to the central halo galaxy. To compare results of different studies we have converted all halo masses to the common definition, $\Mth$ -- the mass within the radius enclosing overdensity of 200 times the critical density of the universe. We omit error bars for most data sets for clarity, but note that they are typically comparable to the symbol size, except at small masses where error in mass in weak lensing is substantial. Note that relations inferred from weak lensing stacking and satellite kinematcis are statistical in nature and provide average halo mass at a given stellar mass. Nevertheless, in some of the studies the authors apply a correction to infer the true $M_{\ast}-M$ relation. Such corrections are, however, uncertain because scatter in the relation is not well known. We therefore also plot abundance matching predictions with scatter for both
$\langle M_\ast\rangle-M_{200}$ and $M_\ast-\langle M_{200}\rangle$ relations to show the potential difference.

The figure shows that galaxies follow a well defined relation from the smallest probed masses to the BCGs in the most massive clusters. In particular, the BCGs in clusters studied in this paper smoothly extend the mean relation derived for smaller mass galaxies via stacked weak lensing analyses and satellite kinematics. The relation derived under the abundance matching approach using the \citet{bernardi_etal13} SMF at the high mass end is in reasonable agreement with direct observational measurements over the entire range of galaxy masses probed by observations. 

\begin{figure}[t]
\begin{center}
\vspace{-1cm}
\includegraphics[scale=0.475]{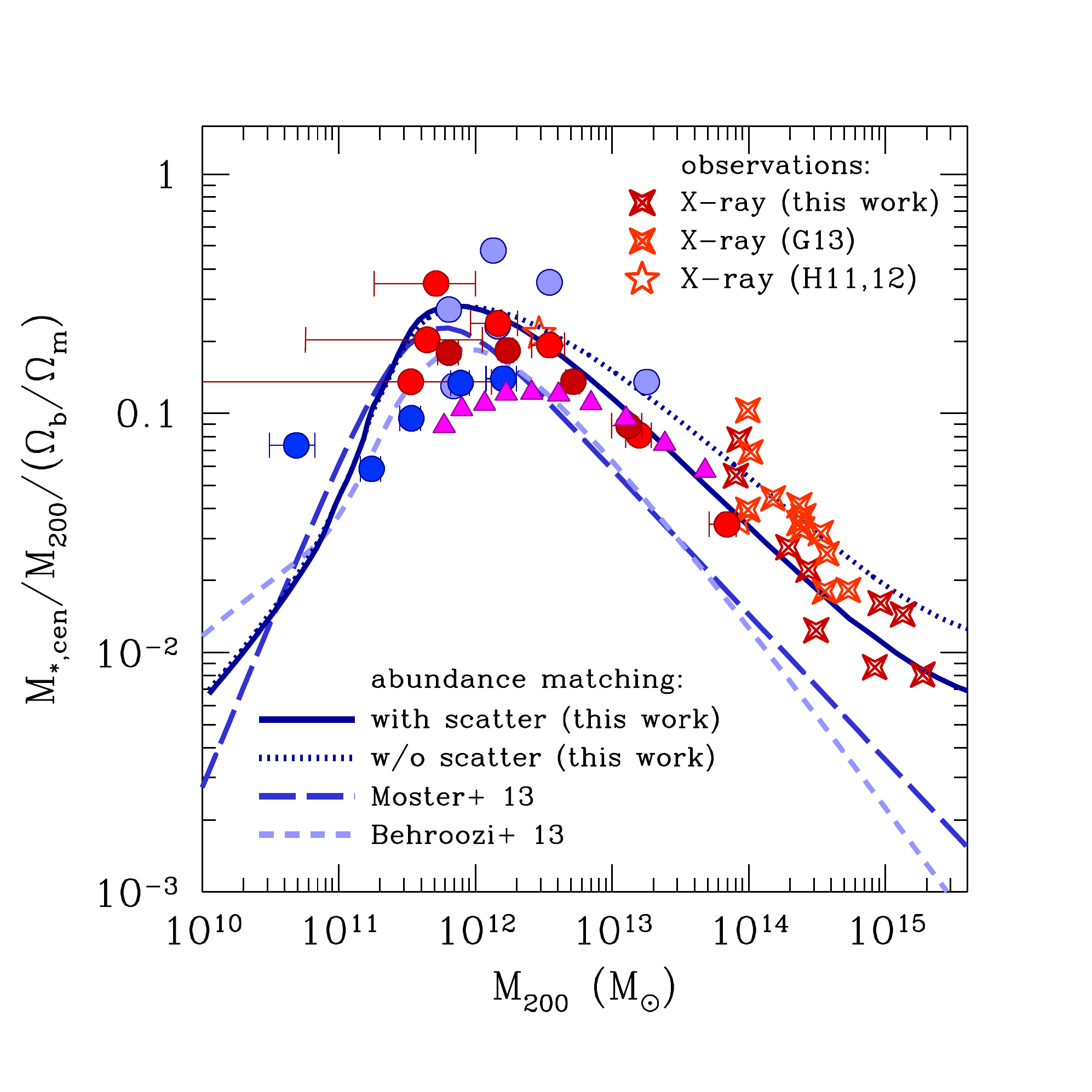}
\vspace{-0.8cm}
\caption{The stellar mass fraction of the central galaxy  in units of the universal baryon fraction within radius $\Rth$ as a function of total halo mass $\Mth$. The stellar masses for clusters analyzed in this study and in G13 are shown by red crosses, while the other symbols are the same as in Figure~\ref{fig:msmh}. The lines show the relations using abundance matching ansatz in this work using the \protect\citet{bernardi_etal13} stellar mass function and previous results by \citet{moster_etal13} and \citet{behroozi_etal13a}.}
\label{fig:fscen}
\end{center}
\end{figure}

 Figure~\ref{fig:fscen} shows the corresponding plot of the stellar mass fraction as a function of halo mass. The stellar mass fraction is normalized by  the universal baryon fraction $\Omega_{\rm b}/\Omega_{\rm m}\approx 0.16$ \citep{hinshaw_etal13}, to gauge more conveniently what fraction of the baryon budget is in stars.
 The figure shows the well known peak in stellar fraction at $M_{200}\approx 10^{12}\ M_{\odot}$ and a steady decrease at both low and high masses.  The stellar mass fraction in central galaxies in $M_{200}\approx 10^{14}\ M_{\odot}$ halos is roughly 5-6 times lower than at the peak. Note that although this decrease is substantial, it is considerably smaller than the decrease implied  by the previously derived $M_\ast-M_{200}$ relations \citep{moster_etal13,behroozi_etal13a}, which predicted stellar mass $\gtrsim 20$ lower than the peak at cluster masses. This shows that suppression of stellar masses in central cluster galaxies is milder than previously thought. 
 
\begin{figure}[t]
\begin{center}
\vspace{-1cm}
\includegraphics[scale=0.475]{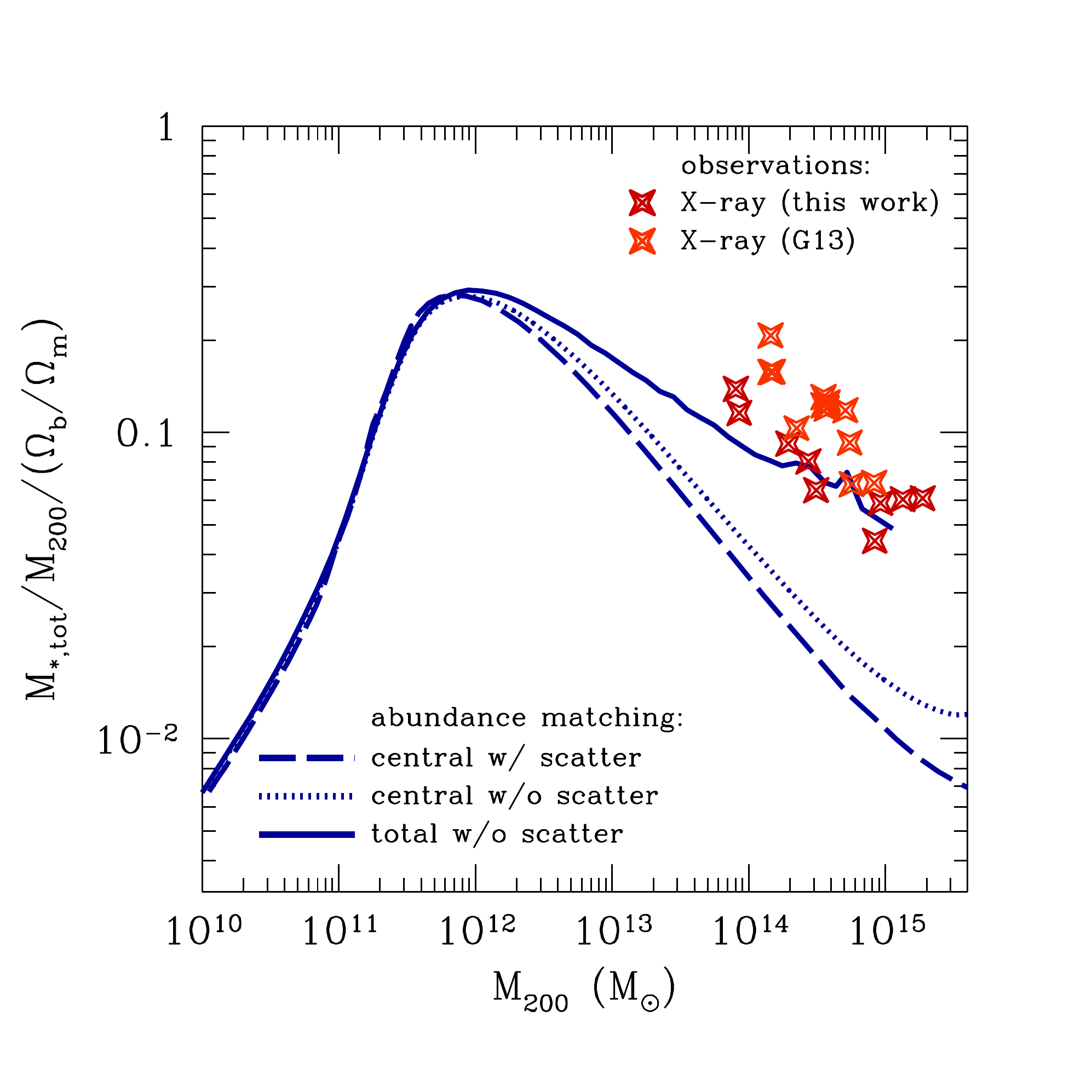}
\vspace{-0.8cm}
\caption{Total stellar fraction (due to the central and all of the satellite galaxies)  within radius $\Rth$ in units of the universal baryon fraction. The crosses show the fractions derived for massive clusters in the cluster sample analyzed in this paper and by G13. The dashed and dotted lines show the abundance matching relations for the central galaxies from Figure~\ref{fig:fscen} for comparison. The solid line shows the expectation fro the total stellar fraction from abundance matching derived using the halos and subhalos in the Bolshoi cosmological simulation, as described in the text.}
\label{fig:fstot}
\end{center}
\end{figure}

The overall efficiency of star formation in cluster halos can be represented by the total stellar mass fraction in all cluster galaxies, including the BCG and satellite galaxies, shown in Figure~\ref{fig:fstot}. This figure shows the total stellar fractions in the observational cluster sample and expected from the abundance matching models. For the latter we use the $M_*-M_{200}$ relation derived using the abundance matching and \citet{bernardi_etal13} SMF to assign stellar masses to halos and subhalos in the Bolshoi simulation, as described above in Section~\ref{sec:msmhtothigh}.

Figure~\ref{fig:fstot} shows that total stellar fractions derived by abundance matching  are in reasonable agreement with observations. In contrast to the mass fraction in the central galaxy, the total stellar fraction is a weak function of halo mass decreasing only by a factor of three when total halo mass changes by more than two orders of magnitude \citep[see also][]{leauthaud_etal12a,leauthaud_etal12b}. This shows that the overall efficiency of star formation per unit mass in massive halos is only a factor of $\sim 3-6$ smaller than in the $M_{200}\sim 10^{12}\ M_{\odot}$, in which star formation efficiency reaches its peak. The key difference between $M\approx 10^{12}\ M_{\odot}$ and cluster halos is that the former form most of their stars in the central galaxy, while a large fraction of stellar mass in cluster halos is contributed by satellite galaxies (see Fig.~\ref{fig:fbcg}). Also, a significant fraction of stellar mass in the central galaxy in clusters is contributed by its extended outer component, while in the MW-sized galaxies stellar halo, which could be considered to be an equivalent of such component, contributes only $\sim 1-2\%$ of stellar mass \citep[e.g.,][]{purcell_etal07}.

\subsection{Effects of IMF}
\label{sec:imf}

Recently, several observational studies have inferred that stellar IMF of early type  galaxies becomes increasingly ``bottom-heavy'' with increasing velocity dispersion and stellar mass. These studies used a variety of techniques, from indirect dynamical constraints on the mass-to-light ratios of stars \citep{grillo_etal09,grillo_gobat10,treu_etal10,auger_etal10,sonnenfeld_etal12,dutton_etal11,dutton_etal12,dutton_etal13,cappellari_etal13b,conroy_etal13} to direct probes of abundance of dwarf stars with $m\approx 0.1-0.2\Msun$ relative to abundance of $m\approx 1\Msun$ stars using unique spectral features of dwarf stars  \citep{vandokkum_conroy10,spiniello_etal12,conroy_etal12}. Currently, observational constraints probe only the inner regions of galaxies ($\lesssim R_e$, where $R_e$ is effective radius of galaxy surface brightness profile) and IMF may vary with galaxy radius. Nevertheless, if we assume that  results for the inner regions are applicable for the entire stellar populations of these galaxies, we can obtain an upper limit on the possible effect of IMF variation on the stellar content of clusters. 

To model effect of such variation, we adopt recent calibration of the trend of the stellar mass-to-light ratio with galaxy velocity dispersion for a sample of compact early type galaxies by \citet[][see their Fig. 3]{conroy_etal13}. The trend is defined with respect to the fiducial Milky Way Chabrier IMF and parametrized as:
\begin{equation}
\log_{10}\left[\frac{(M/L)_*}{(M/L)_{\rm Chabrier}}\right]=a + b\log_{10}\left(\frac{\sigma}{130\ {\rm km\,s^{-1}}}\right),
\end{equation}
where $\sigma$ is velocity dispersion within the SDSS fiber and $a=0.13$ and $b=0.9$ approximate the trend in Figure 10 of \citet{conroy_etal13}. A similar but somewhat weaker trend was also derived recently using dynamical modelling by \citet{cappellari_etal13b}. We adopt a stronger trend of \citet{conroy_etal13} here in the spirit of estimating the upper limit on the effect. We use the relation between the central velocity dispersion and stellar mass for early type galaxies given by eq.~5 of \citet{cappellari_etal13b} and associated parameters to convert velocity dispersion to stellar mass and obtain the corresponding variation of the mass-to-light ratio as a function of stellar mass {\it for early type galaxies}.

\begin{figure}[t]
\begin{center}
\includegraphics[scale=0.475]{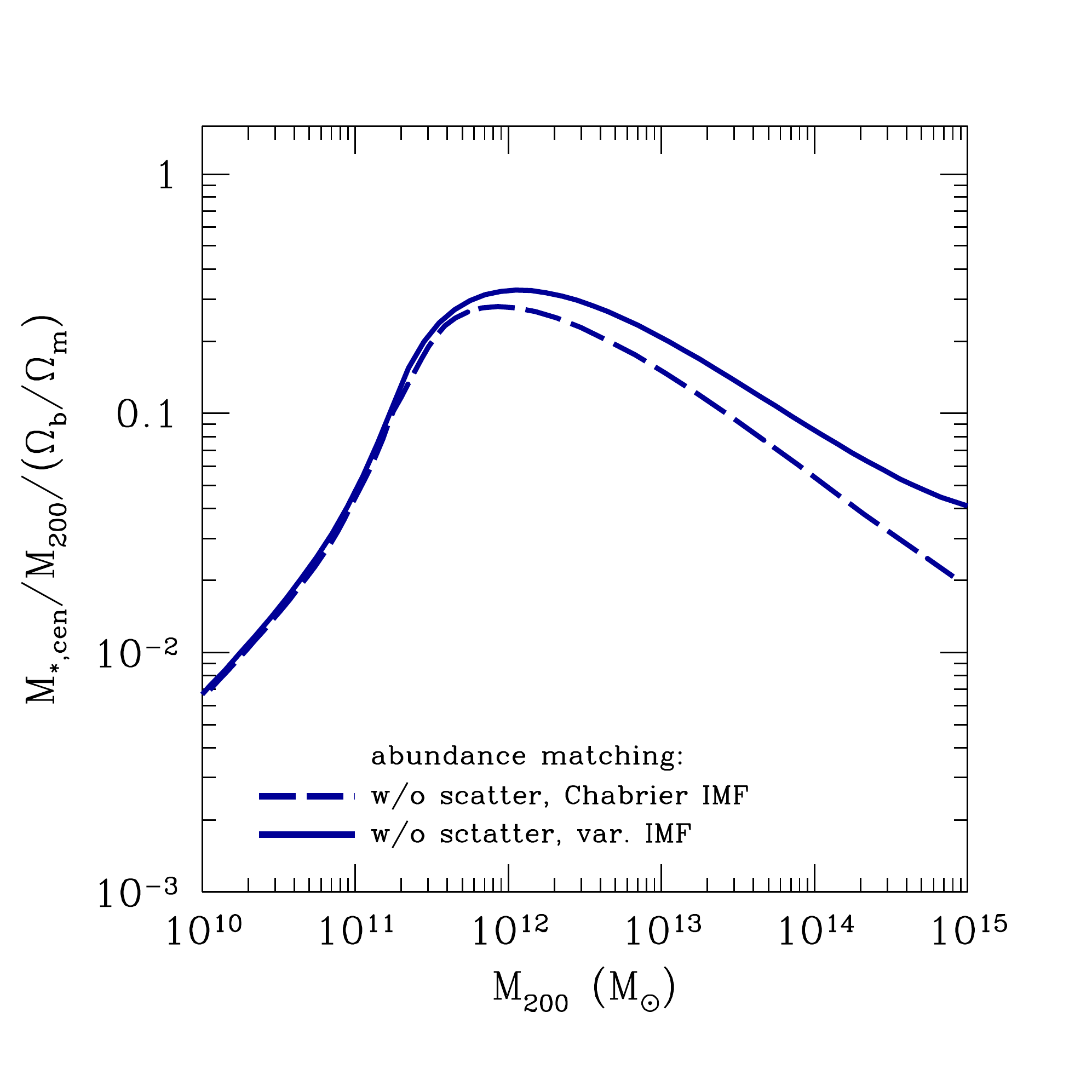}
\vspace{-1cm}
\caption{The stellar mass fraction of the central galaxy  in units of the universal baryon fraction within radius $\Rth$ as a function of total halo mass $\Mth$. The dashed is prediction of the abundance matching model without scatter using the SMF of \protect\citet{bernardi_etal13} with fixed Chabrier IMF. The solid line shows the abundance matching prediction using the SMF, in which stellar masses of early type galaxies have been corrected for the varying IMF, as described in the text. }
\label{fig:fscenvarimf}
\end{center}
\end{figure}

\begin{figure}[t]
\begin{center}
\includegraphics[scale=0.475]{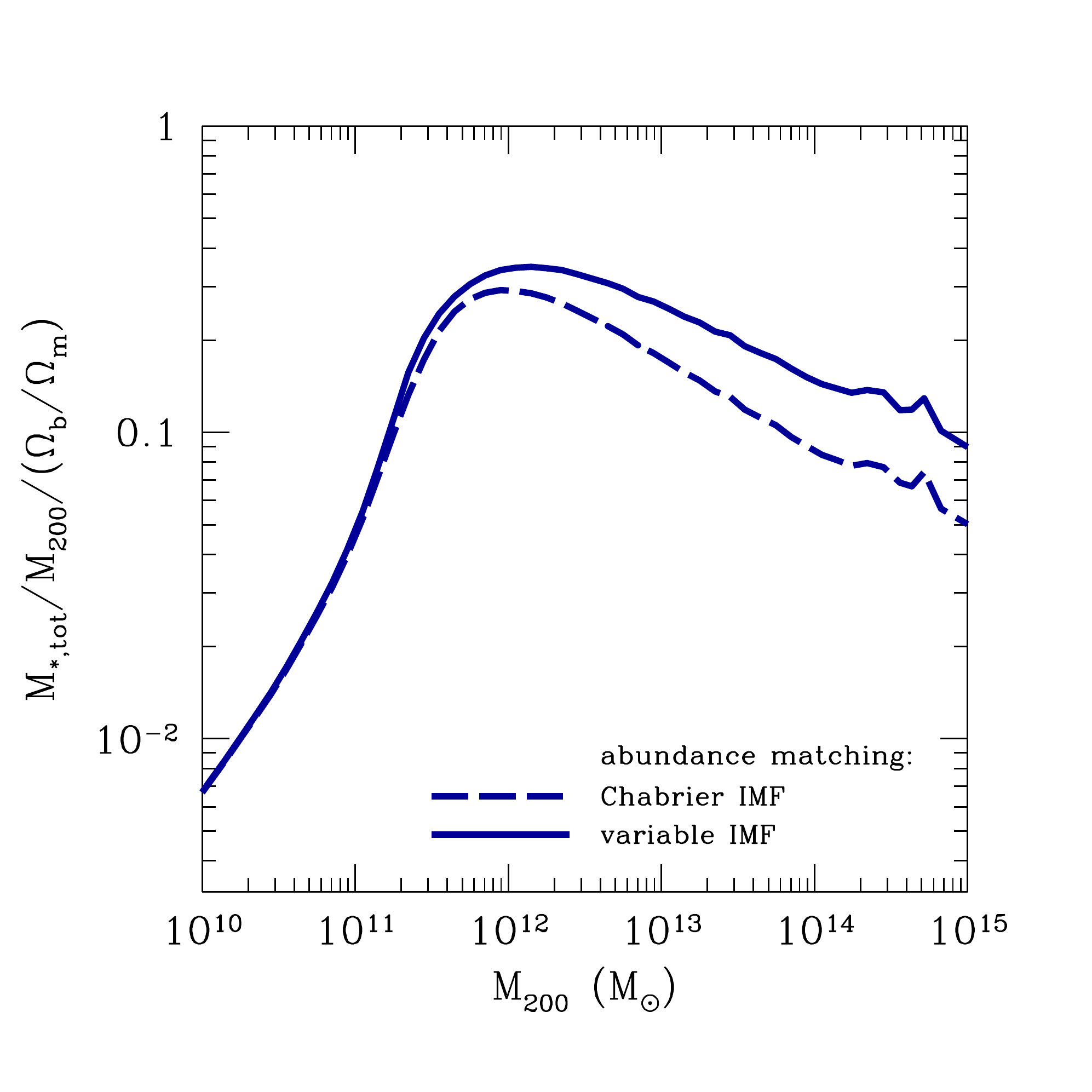}
\vspace{-1cm}
\caption{Total stellar fraction within radius $\Rth$. The dashed line shows prediction of the abundance matching model without scatter using the SMF of \protect\citet{bernardi_etal13} with fixed Chabrier IMF. The solid line shows the abundance matching prediction using the SMF, in which stellar masses of early type galaxies have been corrected for the varying IMF, as described in the text. Because in the latter case IMF becomes progressively more bottom-heavy with increasing galaxy mass, the total stellar fraction exhibits a weaker dependence on halo mass compared to the stellar fraction inferred assuming fixed IMF.}
\label{fig:fstotvarimf}
\end{center}
\end{figure}

To evaluate the effect of IMF variation on the stellar mass function, we use the  parametrizations of the SMFs for late and early type galaxies provided in Table 3 of  \citet{bernardi_etal13}. Given that systematic trend of IMF is deduced for early type galaxies, we only correct the combined SMF of lenticular and elliptical galaxies by the mass-to-light ratio dependence as a function of stellar mass, derived as described above, while leaving the SMF of late type galaxies intact. We then construct the combined SMF as a sum of the early type galaxy SMF corrected for IMF variation and the uncorrected SMF of late type galaxies. We then re-derive the stellar mass--halo mass relation using abundance matching with this new SMF. The results for the fraction of stellar mass in the central galaxy and for the total stellar mass fraction within $\Rth$ are shown in Figures~\ref{fig:fscenvarimf} and \ref{fig:fstotvarimf}. 

As expected, the increase of stellar mass-to-light ratio with increasing stellar mass for early type galaxies results in steeper $M_*-M_h$ relation and, correspondingly, shallower dependence of $f_{\ast,\rm cen}$ on halo mass. The stellar mass fraction of the central galaxies increases by a factor of $\sim 1.5-2$. In this case, for $M_{200}\approx 10^{14}\rm\ M_{\odot}$ halos stellar fraction of the central galaxy is only a factor of $\approx 3-4$ smaller than the peak fraction at $M_{200}=10^{12}\rm\ M_{\odot}$. Likewise, the total stellar fraction within $\Rth$ has increased by a factor of $\approx 1.6-1.7$ in the cluster halos or by $\approx 0.04-0.06$ in absolute values in units of $\Omega_{\rm b}/\Omega_{\rm m}$. This increase is non-negligible for the total baryon census in clusters and can, in principle, explain part of the ``baryon bias'' -- the difference between estimated baryon content of clusters and the universal fraction (see, e.g., G13). It is worth keeping in mind, however, that the effect of IMF variation predicted by our model is an upper limit and the actual effect can be significantly smaller.

The main effect of varying IMF is to make the overall slope of the $f_{\ast,\rm tot}-M_{200}$ relation considerably shallower. In this case, the total stellar fraction in units of the universal fraction varies between the peak value of $\approx 0.35$ at $M_{200}\approx 10^{12}\ \rm M_{\odot}$ and $\approx 0.1$ at $M_{200}\approx 10^{15}\rm\ M_{\odot}$. A similar conclusion was reached in a recent study by \citet{mcgee_etal13}, although these authors have studied models in which IMF variations were even stronger than considered here.

\section{Discussion}
\label{sec:disc}

We have presented measurements of stellar mass for galaxies in nine clusters using SDSS data, using new photometry based on careful modelling of large-scale background and extended surface brightness profiles of massive galaxies.
We show that stellar masses of BCGs derived in our analysis are a factor of $\approx 2-4$ larger than the masses derived from the SDSS cmodel magnitudes, but are close to the stellar masses derived by the photometric analysis used in recent new measurement of the stellar mass function of galaxies by \citet{bernardi_etal13}. 

Combining stellar mass measurements with total masses measured using X-ray Chandra data for these clusters and with similar measurements for 12 clusters from \citet{gonzalez_etal13} allows us to study stellar mass--halo mass relation for the highest mass halos at $z\approx 0$. We show that the relation between BCG stellar mass and total cluster mass has significantly higher normalization and steeper slope than in the previous observational studies (see also discussion in Section \ref{sec:leaucomp} below). Our results thus indicate that stellar and AGN feedback parameters in semi-analytic models and simulations, which were tuned to reproduce previous calibrations of the $M_{\ast}-M$ relation, significantly overestimate the effects of feedback and over suppress star formation in massive halos.  

\begin{figure*}[th]
\vspace{-10cm}
\begin{center}
\hspace{-1.8cm}\includegraphics[scale=0.95]{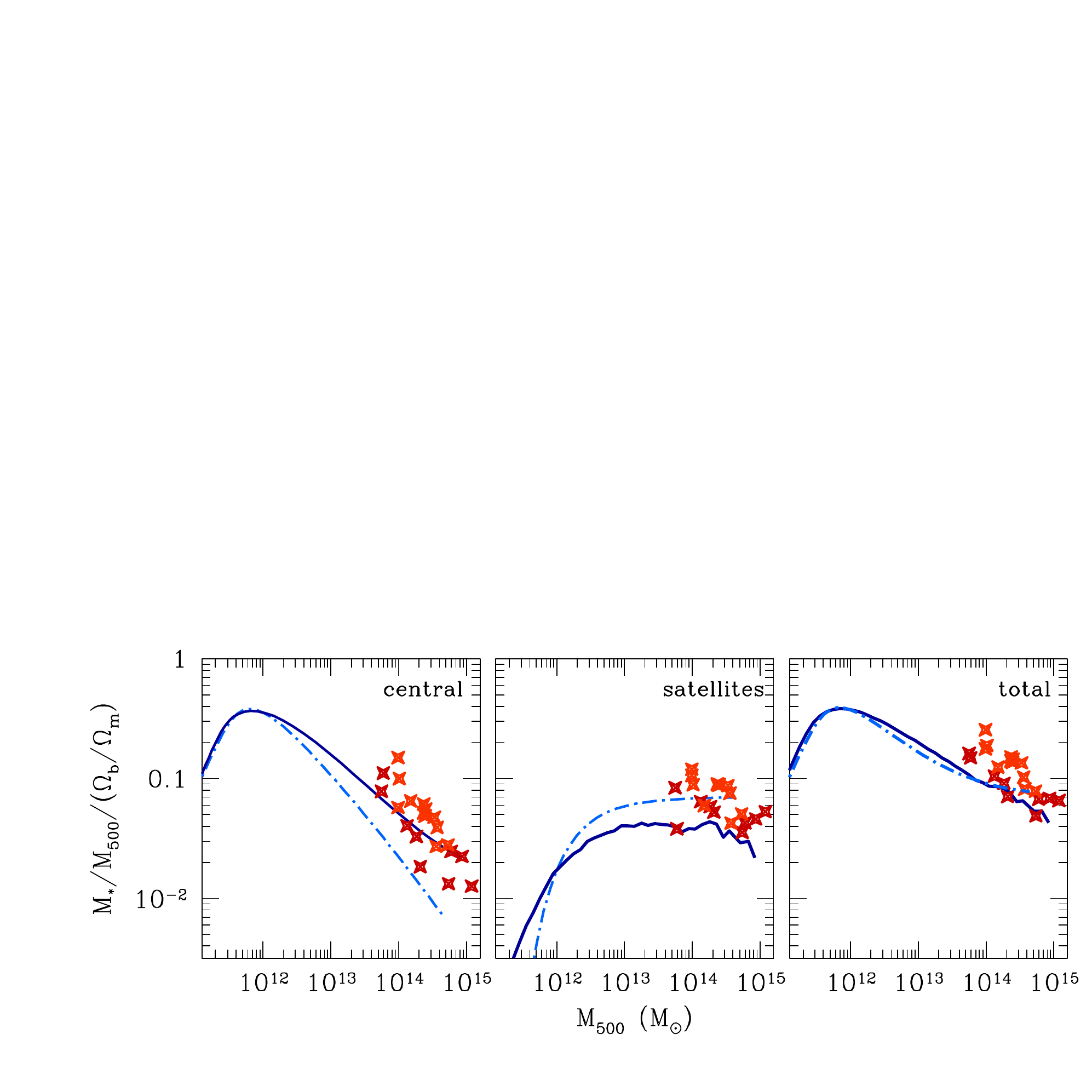}
\vspace{-1cm}
\caption{Stellar mass fraction within radius $\Rfh$ contributed by the central galaxies  (left), satellite galaxies (center), and total stellar fraction of all galaxies (right). The solid line shows prediction of the abundance matching model without scatter using the SMF of \protect\citet{bernardi_etal13} with fixed Chabrier IMF; for the satellite and total contribution we used halo catalog from the Bolshoi simulation to estimate the fractions of stars within $\Rfh$. The dot-dashed lines show the corresponding stellar fractions derived by \protect\citet{leauthaud_etal12b}. The stellar mass fractions for clusters analyzed in this study and in G13 are shown by red crosses (the G13 clusters are shown by light red). The figure shows that the main difference from the results of \protect\citet{leauthaud_etal12b} and direct measurements of stellar fractions in clusters is in the stellar fraction of central galaxies.}
\label{fig:fsleau}
\end{center}
\end{figure*}

Our results also indicate that high mass halos are only modestly less efficient in converting their baryons into stars than halos hosting $L_{\ast}$ galaxies. In particular, the total stellar fraction in cluster halos is a factor of $\sim 3-6$ smaller than the peak fraction in $M\approx 10^{12}\ \rm M_{\odot}$ halos, if the IMF is universal, and could be only $\sim 1.5-3$ times smaller than the peak value, if the IMF in early type galaxies is bottom-heavy, as indicated by recent observational analyses. 

This may seem surprising given that there are many processes that could be invoked to suppress star formation in such massive halos. For example, progenitors of these halos  spend a significant fraction of time in the ``hot accretion'' regime, which is thought to suppress direct accretion of cold gas onto galaxies. Nevertheless, a significant fraction of mass that builds up   group and cluster halos is contributed by $\sim 10^{12}-10^{13}\rm\ M_{\odot}$ halos, i.e. halo masses close to the peak of star formation efficiency. As host halo mass increases, the dynamical friction time for orbital decay of such accreted galaxies increases as well. Thus, although star formation in the BCG galaxy is suppressed, satellites contribute increasingly larger fraction of the total stellar mass fraction in halos of larger mass. 

The satellite galaxies that do merge or get tidally disrupted contribute substantial stellar mass to the outer regions of the BCG galaxies  \citep[e.g.,][]{conroy_etal07}. For example, the model of \citet{behroozi_etal13a} predicts that disrupted satellites deposit up to $\approx 10^{12}\ \rm M_{\odot}$ of stellar mass to the ICL in $M\approx 10^{14}\ \rm M_{\odot}$ clusters, a factor of $\approx 4-5$ larger than the stellar mass of the BCGs in halos of that mass (see their Fig. 9). 
Thus, the total stellar mass of the BCG and ICL predicted in their model is in reasonable agreement with our measurement of the BCG mass that include stellar mass at large radii. This once again illustrates that the discrepancy of the BCG mass between our measurements and their model is  simply due to the fact that their inference is based on observational stellar mass function that do not measure the outer profiles of BCGs properly. The stellar mass of disrupted satellites in their model thus has to be attributed to a separate ICL component rather than to the outer regions of the BCG galaxy. In other words, their model for satellite accretion and disruption actually predicts that the stellar mass associated with the central cluster galaxies is as massive as in our analysis. Our results provide an interesting avenue for testing such models by comparing stellar mass distribution in the observed BCG galaxies and predicted by the model.  

These considerations show that the fact that massive halos have stellar fractions only moderately lower than those of $\sim L_{\ast}$ galaxies, is due primarily to the hierarchical nature of structure formation and high efficiency of star formation in their progenitors that contribute the bulk of the mass.

\subsection{Comparison with previous studies}
\label{sec:leaucomp}

Stellar mass fractions have recently been estimated in a number of studies  \citep[][G13]{lin_etal03,lin_etal12,gonzalez_etal07,giodini_etal09,andreon10,andreon12,zhang_etal10,lagana_etal11,lagana_etal13,leauthaud_etal12a,leauthaud_etal12b,budzynski_etal14}
A detailed comparison between these studies and measurements of the total stellar fraction for the G13 clusters is presented in \S~5.5 of G13 (see their Figure~10).   The main difference between G13 and previous studies is in the inclusion of the mass in the outer regions of BCGs, which is usually missed by the standard survey photometry employed in previous measurements. 
For example, G13 measurements are consistent with most other studies if stellar mass of the central galaxies is estimated within central 50 kpc only. Given that we also include stellar masses at large radii,  our results are in good agreement with G13 and similar conclusion holds for comparison of our results with previous studies.  
 
Total stellar mass fractions derived by \citet{leauthaud_etal12b} using the halo modelling of the COSMOS data are the most discrepant with the results of G13. In particular, the total stellar fraction inferred by \citet{leauthaud_etal12b} is almost constant (at $\approx 0.08-0.09$ in units $\Omega_{\rm b}/\Omega_{\rm m}$ for the Chabrier IMF) over the mass interval probed by cluster data. A similar conclusion was recently  reached by \citet{budzynski_etal14}, who used optically-selected SDSS group and cluster catalog of \citet{budzynski_etal12} and measured stellar fractions both using SDSS cmodel photometry of cluster galaxies and including all of the light measured by stacking cluster images. For Chabrier IMF they derive an approximately constant total stellar fraction of 0.08 in units of $\Omega_{\rm b}/\Omega_{\rm m}$ for the interval $M_{500}\approx 10^{14}-10^{15}\rm\ M_{\odot}$. The total stellar fractions of G13 and our clusters, however,  systematically decrease with increasing cluster mass. 

We compare stellar fractions measured for our clusters along with measurements of G13 to the stellar fractions contributed by the central galaxy, satellite galaxies, and the total fraction (central$+$satellites) with the HOD results of \citet{leauthaud_etal12b} in Figure~\ref{fig:fsleau}. The figure also shows results of the abundance matching analyses presented in this study and results for central galaxies from \citet{moster_etal13} and \citet{behroozi_etal13a}. The left panel of Figure~\ref{fig:fsleau} shows that the stellar fraction of central galaxies derived by \citet{leauthaud_etal12b} is close to our results at $M_{500}\lesssim 10^{13}\rm\ M_{\odot}$, but is significantly below direct measurements and our abundance matching results for $M_{500}\gtrsim 10^{14}\rm\ M_{\odot}$. At high masses their results are close to the AM results of \citet{moster_etal13} and \citet{behroozi_etal13a}. As we discussed above, these analyses use SMFs in which stellar masses of central galaxies were significantly underestimated \citep{bernardi_etal13} and therefore underestimate stellar fractions of central galaxies in the high mass halos. 

Comparison of stellar fractions contributed by satellite galaxies in the middle panel shows that there is reasonable agreement between results of \citet{leauthaud_etal12b} and direct measurements in clusters. The main source of the difference in the total stellar fractions at $M_{500}\approx 10^{14}\rm \ M_{\odot}$, apparent in the right panel of Figure~\ref{fig:fsleau}, is thus in the masses of BCG galaxies. We find that these galaxies contribute a significant fraction of the total stellar mass. Decrease of the BCG stellar fraction with increasing halo mass is thus the source of the corresponding decrease of the total stellar fraction with increasing halo mass.

The reasons for a similar discrepancy with results of \citet{budzynski_etal14} are less clear as these authors do include contribution of the extended component in their analysis of stacked images. Surprisingly, however, they find that for $M_{500}\approx 10^{14}\rm\ M_{\odot}$ clusters the total stellar fractions obtained via such image stacks are consistent with fractions estimated using SDSS model galaxy magnitudes, which miss the contribution of the extended component, as shown by \citet{bernardi_etal13} and in our Section~\ref{sec:lcomp}. However, the photometry errors should be smaller for the more distant systems in the \citet{budzynski_etal14} sample. 
A possible other source of bias is a large scatter in the richness mass relation and associated contamination of the lower mass end of their sample by low mass systems. For such systems masses are overestimated and stellar fractions are therefore underestimated.

\subsection{Implications for galaxy formation models and feedback}
\label{sec:impl}

Stellar content of group and cluster halos and the overall stellar mass function of galaxies are frequently used as a diagnostic of feedback prescriptions in models of galaxy formation. For example, stellar mass--halo mass relation is often used to tune parameters of semi-analytic models and phenomenological recipes in galaxy formation simulations. We show that normalization of this relation is a factor of $\approx 2-4$ higher than derived in previous studies and suppression of star formation in massive halos by AGN and stellar feedback is therefore considerably weaker than previously thought. Prescriptions for such feedback in semi-analytic models thus need to be revised. 

Cluster formation simulations generically predict that total stellar fraction is only a weak function of cluster mass \citep[e.g.,][]{kravtsov_etal05,puchwein_etal10,mccarthy_etal11,planelles_etal13,martizzi_etal13}, i.e. total stellar mass scales almost linearly with cluster mass. Varying physical prescriptions for star formation changes the overall value of the total stellar fraction, but not its (in)dependence on total mass \citep[e.g.,][]{puchwein_etal10,mccarthy_etal11,planelles_etal13}. In contrast, stellar fraction in observed clusters considered in our study and by G13 show a systematic decrease of the total stellar fraction with increasing cluster mass. The reason for the difference in this case appears to be a nearly linear scaling of stellar mass of both the BCG and satellite galaxies with the total cluster mass  \citep[e.g.,][]{puchwein_etal10,planelles_etal13,martizzi_etal13}. 

\begin{figure}[t]
\begin{center}
\includegraphics[scale=0.475]{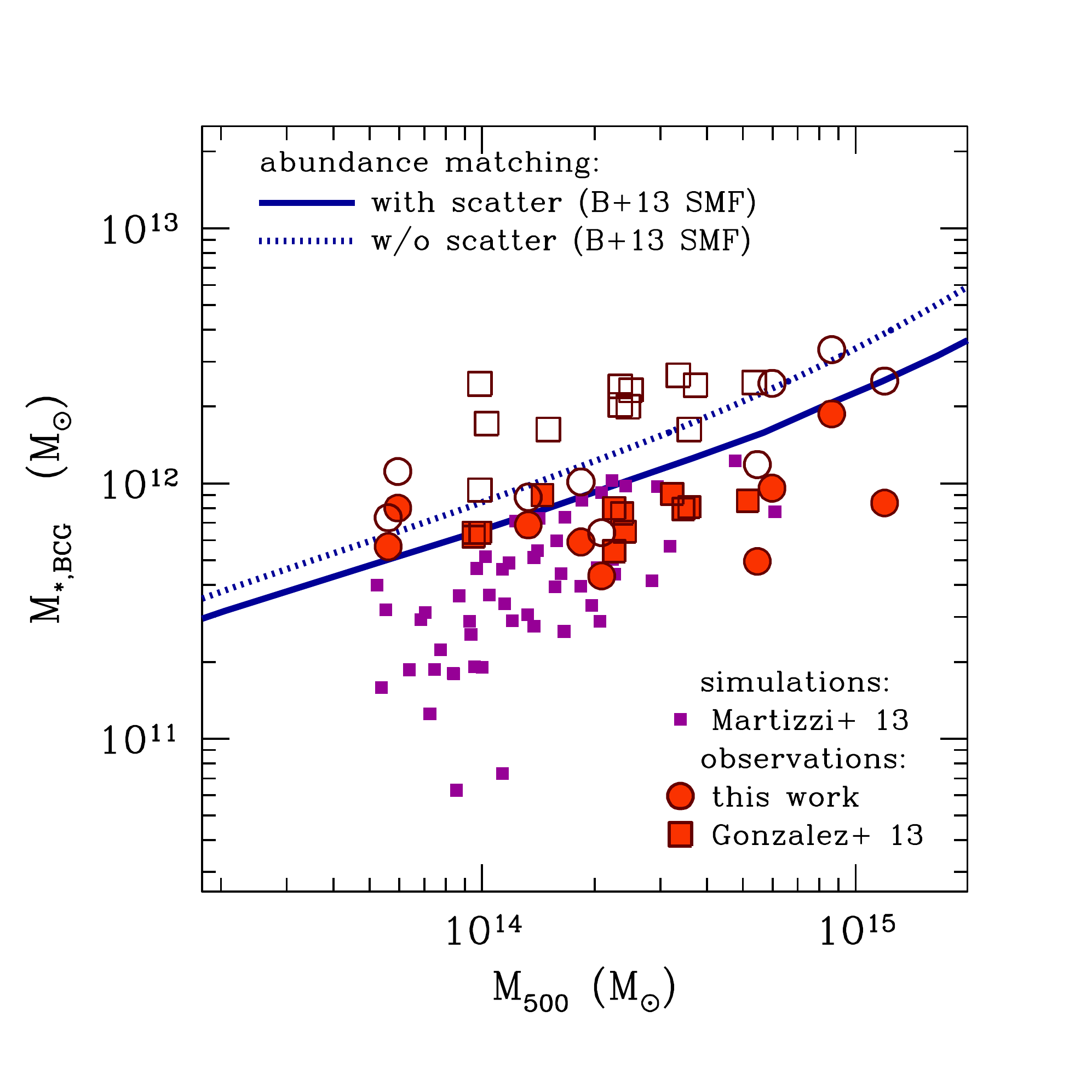}
\vspace{-1cm}
\caption{Stellar masses of the BCG galaxies  versus total halo mass, $\Mfh$, in clusters presented in this study and the study of G13 compared to predictions of simulations by \protect\citet{martizzi_etal13}. For our clusters we define stellar masses in exactly the same way as \protect\citet{martizzi_etal13}: i.e., by counting all of the stellar mass in regions around the center where 3D stellar density is $\rho_*>2.5\times 10^5\,\rm M_{\odot}\,kpc^{-2}$. These stellar masses are shown as red solid circles. For the G13 clusters we show (red solid squares) stellar masses measured within central 50 kpc, which approximately corresponds to an average radius defined by the above stellar density threshold condition. Corresponding open circles and squares show the total stellar masses of central galaxies and their extended halos presented in previous figures. Magenta points show predictions of simulations with AGN feedback by \protect\citet{martizzi_etal13}. 
The solid and dotted lines show relation derived using the abundance matching approach with and without scatter using the SMF of \protect\citet{bernardi_etal13}.  } 
\label{fig:msmhmartizzi}
\end{center}
\end{figure}

The discrepancy indicates that feedback and other relevant physical processes included in simulations need further improvement. At the same time, some differences may arise simply due to differences in how stellar mass is defined and measured. Therefore, care should be taken to measure stellar mass in simulation analysis mimicking observational measurement as closely as possible.
The importance of mass definition is illustrated in Figure~\ref{fig:msmhmartizzi}, where we show comparison of stellar masses of the BCGs in observed clusters in our sample and recent simulations by \citet{martizzi_etal13}. \citet{martizzi_etal13} define stellar mass associated with the central galaxy using three-dimensional stellar density profile and counting all of the stars within radius where local stellar density is $>2.5\times 10^{5}\rm\,M_{\odot}\,kpc^{-3}$. This density threshold corresponds to physical radii of $\approx 40-60$~kpc (D. Martizzi, priv. communication). The figure shows comparison of the stellar masses of simulated BCGs defined in this way with the total stellar mass of observed BCGs and their stellar mass within the central 50 kpc. For the latter definition, simulations underestimate stellar masses at $M_{500}\approx 10^{14}\rm\ M_{\odot}$, while at larger masses they match stellar masses. However, the difference between open and solid symbols in Figure~\ref{fig:msmhmartizzi} shows that a large fraction of stellar mass in observed galaxies is at radii $>50$ kpc and detailed comparisons with simulations should also include comparisons of stellar mass defined within larger radii.  We advocate using masses measured within specific apertures for such comparisons and provide such measurements for our BCGs for a range of apertures in Table~\ref{tab:sdssclalt} in the Appendix B. 

\section{Conclusions}
\label{sec:conc}

In this study we investigated stellar mass--halo mass relation and corresponding efficiency of star formation for high mass halos using a sample of galaxy clusters with careful measurements of stellar and total masses.   
Our main results and conclusions can be summarized as follows.
\begin{itemize}

\item[1.]  Stellar surface density profiles of the BCGs in the nine clusters analyzed in this study are well described by  three S\`ersic components and extend smoothly out to $\gtrsim 100-300$ kpc. The cumulative stellar mass of the BCG galaxies continues to grow at the last reliably measured radius in all of the BCGs.

\item[2.] We show that the stellar surface density profiles of BCGs in the inner regions are similar to the stellar surface density profiles of lower mass elliptical galaxies and to the outer regions of late type galaxies. Sizes of BCG galaxies correlate with the virial radius of host cluster halo and smoothly extend the corresponding correlation for smaller mass galaxies.

\item[3.] We find that stellar mass of the BCGs correlates with the total cluster mass as  $M_{*,\rm BCG}\propto M_{500}^{\alpha_{\rm BCG}}$ with the best fit slope is $\alpha_{\rm BCG}\approx 0.4\pm 0.1$ and scatter of $\approx 0.17$ dex. For the stellar mass in satellite galaxies $M_{*,\rm sat}\propto M_{500}^{\alpha_{\rm sat}}$, the best fit slope is $\alpha_{\rm sat}\approx 0.8\pm 0.1$ and scatter is $\approx 0.1$ dex.  The total stellar mass exhibits power law scaling with $\alpha_{\rm tot}\approx 0.6\pm 0.1$ and scatter of $0.1$ dex.  The overall efficiency of baryon conversion into stars in individual clusters of a given mass thus shows rather little variation.  

\item[4.] We show that the total stellar fractions in clusters are only a factor of $\sim 3-5$ lower than the peak stellar fraction reached in $M\approx 10^{12}\rm\ M_{\odot}$ halos, if the IMF is universal. The difference is only a factor of $\sim 1.5-3$ if the IMF becomes progressively more bottom heavy with increasing mass in early type galaxies, as indicated by several recent observational analyses.  This means that the overall efficiency of star formation in massive halos is only moderately suppressed compared to $L_{\ast}$ galaxies. 

\item[5.] We show that stellar mass--halo mass relations derived using abundance matching and reported in recent studies underestimate masses of the BCG galaxies in clusters by a factor of $\sim 2-4$. We argue that this is because these abundance matching analyses used stellar mass functions based on photometry that does not properly account for the outer surface brightness profiles of massive galaxies. We show that $M_{\ast}-M$ relation derived using abundance matching with the SMF based on photometry which corrects this problem is in a much better agreement with our derived relation. 

\item[6.] The new $M_{\ast}-M$ relation we derive is significantly steeper than the previously derived relations with significantly larger stellar masses at a fixed halo mass for halos with $M\gtrsim 10^{12}\rm\ M_{\odot}$. This new relation indicates that suppression of star formation by the AGN feedback and other processes in massive halos is weaker than previously thought. Implementations of these feedback
processes in semi-analytic models and simulations thus needs to be re-evaluated. 

\end{itemize}

\acknowledgements
We are grateful to Anthony Gonzalez for sending us advanced copy of his submitted paper and useful discussions on stellar mass measurements, and to Mariangela Bernardi and Alan Meert for useful discussions and communicating results of their S\'ersic fits and new calibration of the stellar mass function before publication. We are also grateful to Davide Martizzi for sending us their simulation results in electronic format. AK and AV would like to thank Kavli Institute for Theoretical Physics at UC Santa Barbara for hospitality during the
workshop ``Galaxy Clusters: the Crossroads of Astrophysics and
Cosmology'' (2011), where discussions that led to this work have been initiated. We would also like to thank O. Gnedin, S. Genel, A. Gonzalez, A. Hearin, R. Skibba for comments on the manuscript. 
AK  was supported via NSF grant OCI-0904482, by NASA ATP grant NNH12ZDA001N, and  by the Kavli Institute for Cosmological Physics at the University of Chicago through grants NSF PHY-0551142 and PHY-1125897 and an endowment from the Kavli Foundation and its founder Fred Kavli. AV was supported by NASA Contract NAS8-03060 and Chandra X-ray observatory grant GO2-13142X. AM acknowledges support through Russian Foundation for Fundamental Research (grants 12-02-01358-а and 13-02-01464-a), the Presidential program supporting leading scientific schools (project NSH-6137.2014.2), the Russian Academy of Sciences programs ``Origin, Structure, and Evolution of Objects of the Universe'' (P20) and ``Extended objects in the Universe'' (OFN-16). Special thanks to Predoctoral Fellowship Program at the Harvard-Smithsonian Center for Astrophysics for financial support and provided excellent research opportunities and experience.


\bibliographystyle{apj}
\bibliography{review,imf,xray}

\appendix 
\section{Stellar mass--halo mass relation from abundance matching}
\label{sec:amapp}

To derive stellar mass-halo mass relation using the abundance matching (AM) ansatz, which was shown to work well in reproducing the observed luminosity dependence of galaxy clustering \citep{kravtsov_etal04,tasitsiomi_etal04,conroy_etal06,reddick_etal13}
and other statistics  \citep{vale_ostriker04,vale_ostriker06,behroozi_etal10,behroozi_etal13b,guo_etal10,moster_etal13,hearin_etal13a}.
Specifically, we use 
  \citet{tinker_etal08} calibration of the halo mass function for $\Mtwoh$ and $\Mfh$, which was calibrated using host halos only. To account for subhalos, we correct the host mass function by subhalo fraction, $f_{\rm sub}(>M)=[n_{\rm tot}(>M)-n_{\rm host}(>M)]/n_{\rm host}(>M)$, to get $n_{\rm tot}(>M)$ -- the mass function that includes both hosts and subhalos. The latter was calculated using current  masses for hosts and corresponding masses at the accretion epoch for subhalos using $z=0$ halo catalog halo catalog of \citet{behroozi_etal13c} derived from the Bolshoi simulation \citep{klypin_etal11} of $(250h^{-1}\rm Mpc)^3$ volume in the concordance cosmology adopted in this study. The subhalo fraction in the Bolshoi simulation is parametrized as $f_{\rm sub}={\rm min}[0.35,0.085(15-\log_{10}\Mtwoh)]$. The halo mass function derived from the Bolshoi simulation agrees within $5\%$ with the \citet{tinker_etal08} parametrization, but the latter is more accurate at the highest halo masses. 

We combine two recent calibrations of the SMF by \citet{papastergis_etal12} and \citet{baldry_etal12} at small masses and \citet{bernardi_etal13} at large $M_*$. 
We approximate the SMF measured by \citet{papastergis_etal12} and \citet{baldry_etal12} using double Schechter form given by eq.~6 of \citet{baldry_etal12} with the following parameters:  $\log_{10}(M_*/M_{\odot})=10.66$, $\phi_1^{\ast
}=3.96\times 10^{-3}\ \rm Mpc^{-3}$, $\alpha_1=-0.35$, $\phi_2^{\ast}=6.9\times 10^{-4}\ \rm Mpc^{-3}$, $\alpha_2=-1.57$.
These parameters are in general agreement with the best fit parameters derived for the local stellar mass function by \citet{baldry_etal12}. The SMF of \citet[][]{bernardi_etal13} is also parametrized by a double Schechter function (see Eq. 1 of their paper) with parameters for the Sersic fits to galaxy surface brightness profiles (second line in the SMF section of their Table 1): $\phi_*=1.04\times 10^{-2}\ \rm Mpc^{-3}$, $M_*=9.4\times 10^7\rm\  M_{\odot}$, $\alpha=1.665$, $\beta=0.255$, $\phi_\gamma=6.75\times 10^{-3}\rm\ Mpc^{-3}$, $M_\gamma=2.7031\times 10^9\rm\ M_{\odot}$, $\gamma=0.296$. It is important to note that the SMF of \citet{bernardi_etal13} was estimated using photometry in which particular care was chosen to estimate background for extended massive galaxies. As shown by \citet{bernardi_etal13} this results in a significant enhancement of stellar masses for galaxies of $M_*\gtrsim 10^{11}\ \rm M_{\odot}$. I refer readers to the original papers for further details on how the stellar mass functions were estimated. 

\begin{figure}[t]
\begin{center}
\includegraphics[scale=0.475]{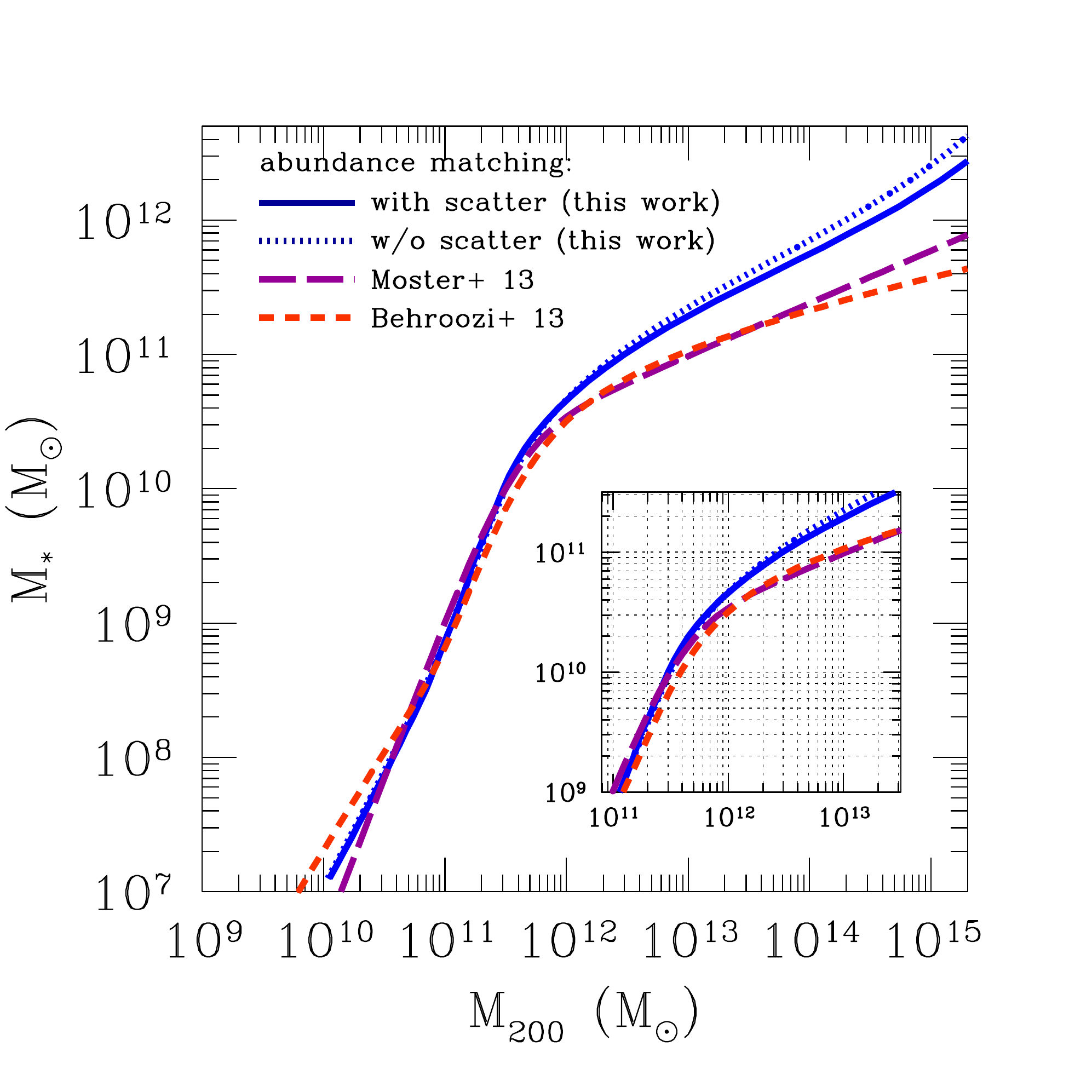}
\caption{Comparison of the $M_*-M$ relations derived using abundance matching ansatz. Solid line shows relation derived in this study assuming a constant scatter of $\log_{10}M_*$ of 0.2 dex at fixed $M_{200}$, while the dotted line shows the corresponding relation for zero scatter. The short and long dashed curves show the relation derived by \citet{moster_etal13} and \citet{behroozi_etal13a}, respectively. The inset shows zoom-in on the region around $M_{200}=10^{12}\rm\ M_{\odot}$.}
\label{fig:msmhcomp}
\end{center}
\end{figure}

We use these two calibrations to construct a combined stellar mass function, $n(M_*)={\rm max}[n_{\rm B12},n_{\rm B13}]$, that spans from $M_*\approx 10^7\rm \Msun$ to $M_*\approx 10^{12}\ \Msun$. Both stellar mass functions assume \citet{chabrier03} IMF to estimate stellar masses of galaxies. The SMF is defined by the SMF of \citet{bernardi_etal13} at $M_*>3\times 10^9\rm\ M_{\odot}$. Thus, the abundance matching results presented in this paper are determined primarily by this SMF. 

We present results both for $M_*-M$ relation obtained from straightforward abundance matching, $n(>M_*)=n(>M)$, without accounting for any scatter between $M_*$ and $M$, and relation in which we assume a fixed scatter in $M_*$ at a fixed $M$ described by the log-normal pdf, $p(M_*\vert M)$:
\begin{equation}
p(M_*\vert M)dM_*=\frac{\log_{10}e}{\sqrt{2\pi}\sigma_{\log_{10}M_*}}\exp\left[-\frac{(\log_{10}M_*-\langle \log_{10}M_*\rangle(M))^2}{2\sigma^2_{\log_{10}M_*}}\right]\frac{dM_*}{M_*}
\end{equation} 
For the case with scatter the $M_*-M$ relation obtained using equation 
\begin{equation}
n(M_{\ast})=\int^{\infty}_0p(M_*\vert M)n(M)dM,
\end{equation}
where we approximate relation between  $\langle\log_{10}M_*\rangle(M)$ and $\log_{10}M$ as linear locally $\langle\log_{10}M_*\rangle=a+b\log_{10}M$ and obtain values of $a_i(M_i)$,$b_i(M_i)$ for a grid of $M_i$ by minimization of the difference: $n(M_{\ast})-\int^{\infty}_0p(M_*\vert M)n(M)dM$.

Figure~\ref{fig:msmhcomp} shows comparison of the $M_*-M_{200}$ results obtained as described above with the relations derived by \citet{moster_etal13} and \citet{behroozi_etal13a} using abundance matching ansatz. Clearly, our relation is much steeper at $M_{200}\gtrsim 10^{12}\rm\ M_{\odot}$. On cluster mass scale ($M_{200}\gtrsim 10^{14}\rm\ M_{\odot}$), the stellar masses predicted by our relation are a factor of $\gtrsim 3$ larger than those predicted by M13 and B13 relations. As we show in \S~\ref{sec:mscmh} our relation is in much better agreement with direct measurements of the $M_*-M$ relation at these masses. The main reason for the differences is that we used SMF of \citet{bernardi_etal13}, which is much shallower at larger $M_*$ due to increasingly larger stellar masses measured for massive galaxies with their improved photometry analysis method. 

Although differences in the $M_*-M_{200}$ relations shown in Fig.~\ref{fig:msmhcomp} are smaller at smaller masses, they are still substantial even at $M_{200}\approx 10^{12}\ \rm M_{\odot}$, as can be seen in the inset showing zoom-in of the region around this mass. For example, for the Milky Way stellar mass of $M_*\approx 5\times 10^{10}\rm\ M_{\odot}$ the corresponding halo mass is $M_{200}\approx 1.3\times 10^{12}\rm\ M_{\odot}$ from our relation, but is $M_{200}\gtrsim 2\times 10^{12}\rm\ M_{\odot}$, but for stellar mass of M31, $M_*\approx 9\times 10^{10}\rm\ M_{\odot}$, corresponds to $M_{200}\approx 3\times 10^{12}\rm\ M_{\odot}$ according to our relation and to $\approx 6\times 10^{12}\rm\ M_{\odot}$ and $\approx 8\times 10^{12}\rm\ M_{\odot}$ according to relations of M13 and B13. 

For reference, we provide best fit parameters for parametrized form of the $M_*-M$ relation based on the abundance matching with the \citet{bernardi_etal13} SMF and assuming the WMAP cosmology of the Bolshoi simulation  in Table~\ref{tab:msmhfit}. We adopt the five-parameter parametrization of \citet[][see their eq. 3]{behroozi_etal13a}:
\begin{eqnarray}
\log_{10}M_{\ast}&=&\log_{10}\left(\epsilon M_1\right) + f\left[\log_{10}\left(\frac{M}{M_1}\right)\right]-f(0)\\
f(x)&\equiv&-\log_{10}\left(10^{\alpha x}+1\right)+\delta\frac{[\log_{10}(1+\exp(x))]^{\gamma}}{1+\exp(10^{-x})},
\end{eqnarray}
where all masses are in $M_{\odot}$. We profide fits for relation with and without scatter. For the former, we adopt a constant scatter in $\log_{10}M_{\ast}$ at a fixed $M$ of 0.2 dex. 
The fits approximate the actual relations obtained from abundance matching to $\lesssim 0.05$~dex for halo masses in the range $10^{10}\ M_{\odot}<M<10^{15}\rm\ M_{\odot}$. Given the systematic errors in SMF and stellar mass measurements are expected to be $\gtrsim 0.1$~dex, this accuracy should be sufficient for most purposes. For higher accuracy the actual abundance matching calculation should be carried out. 

\begin{table*}[t]
\begin{center}
\caption{Parameters of best fit $M_{\ast}-M$ parametrization at $z\lesssim 0.1$}
\label{tab:msmhfit}
\begin{tabular}{lccccc}
\hline\hline\\
 mass definition         &  $\log_{10}(M_1/M_{\odot})$ &   $\log_{10}\epsilon$  & $\alpha$ & $\delta$ & $\gamma$\\\\\hline\\
 $M_{200c}$ (no scatter)       & 11.39 & -1.618 & 1.795 &    4.345 & 0.619\\
 $M_{200c}$ (with scatter)     & 11.35 & -1.642 & 1.779 &    4.394 & 0.547\\
 $M_{500c}$ (no scatter)       & 11.32 & -1.527 & 1.856 &    4.376 & 0.644\\
 $M_{500c}$ (with scatter)     & 11.28 & -1.556 & 1.835 &    4.437 & 0.567\\
 $M_{\rm 200m}$ (no scatter)   & 11.45 & -1.702 & 1.736 &    4.273 & 0.613\\
 $M_{\rm 200m}$ (with scatter) & 11.41 & -1.720 & 1.727 &    4.305 & 0.544\\
 $M_{\rm vir}$ (no scatter)    & 11.43 & -1.663 & 1.750 &    4.290 & 0.595\\
 $M_{\rm vir}$ (with scatter)  & 11.39  & -1.685 & 1.740 &    4.335 & 0.531\\

  \\
\hline
\end{tabular}
\end{center}
\end{table*}

\section{Alternative BCG stellar mass measurements}
\label{sec:altmass}

In many recent studies comparing results of simulations with BCG masses stellar content of galaxies 
attempts are made to model the BCG and its outer regions (the ICL) separately and to mimic the corresponding separation of these components in observational analyses. 
As we discussed above, it is difficult to separate the inner and outer components of BCGs reliably in practice. At the same time, a number of different definitions is employed in the literature which complicates comparisons of observations and models. We advocate the use of total stellar mass within a given well-defined radii for such comparisons and present stellar masses measured for a range of radii for the BCG galaxies in our nine clusters in Table~\ref{tab:sdssclalt}. The typical errors for these masses due to uncertainties of background are $\approx 5-10\%$ for the largest radii and are smaller at smaller radii. Given the small magnitude of these errors we do not quote them in the table. The overall error of stellar masses is dominated by systematic error in conversion of luminosities into stellar mass, which can be as large as $0.1-0.2$ dex \citep[e.g.,][]{conroy_etal13}.

\begin{table*}[t]``
\begin{center}
\caption{Alternative stellar mass measurements for BCG of the nine SDSS clusters}
\label{tab:sdssclalt}
\begin{tabular}{cccccccc}
\hline\hline\\
 Cluster  &  $R_{500}$ &   $M_*(<30\rm\, kpc)$  & $M_*(<50\rm\, kpc)$ &  $M_*(<70\rm\, kpc)$ &  $M_*(<0.05 R_{500})$ &  $M_*(<0.1 R_{500})$  &  $M_*(<0.25 R_{500})$\\
    & kpc   &           \multicolumn{6}{c}{$10^{11}\rm\ M_{\odot}$}\\
\\\hline\\
    A2142 & $1539$ &  $5.18$ &   $7.61$ &   $9.61$ &   $10.21$ &   $14.63$ &   $21.69$ \\  
    A2029 & $1387$ &  $10.44$ &   $15.25$ &   $18.84$ &   $18.74$ &   $25.35$ &   $34.09$ \\ 
      A85 & $1235$ &  $7.12$ &   $9.47$ &   $10.91$ &   $10.38$ &   $13.30$ &   $21.76$ \\ 
    A1795 & $1196$ &  $3.85$ &   $5.29$ &   $6.19$ &   $5.76$ &   $8.14$ &   $12.70$ \\ 
    MKW3s & $ 873$ &  $3.67$ &   $4.78$ &   $5.56$ &   $4.48$ &   $6.11$ &   $7.80$ \\ 
    A2052 & $ 840$ &  $4.35$ &   $6.05$ &   $7.16$ &   $5.45$ &   $7.68$ &   $10.40$ \\ 
    A1991 & $ 748$ &  $4.71$ &   $6.75$ &   $8.20$ &   $5.55$ &   $8.48$ &   $10.84$ \\ 
    MKW4  & $ 568$ &  $4.59$ &   $5.99$ &   $6.87$ &   $4.45$ &   $6.33$ &   $8.48$ \\ 
  RXJ1159$+$5531 & $ 568$ &  $6.76$ &   $8.30$ &   $9.12$ &   $6.58$ &   $8.63$ &   $10.38$ \\ 
  \\
\hline
\end{tabular}
\end{center}
\end{table*}

\end{document}